\documentclass[aos,preprint]{imsart}

\usepackage{amsthm,amsmath,natbib,color}
\usepackage{amssymb}
\usepackage{latexsym}
\usepackage{float}
\usepackage{graphicx}
\usepackage{amsfonts}
\usepackage{bbm}
\usepackage{longtable}
\usepackage{booktabs}
\RequirePackage[colorlinks,citecolor=blue,urlcolor=blue]{hyperref}

\arxiv{}

\startlocaldefs

\def\ds{\displaystyle}

\def\R{\mathbb{R}}

\def\X{\mathcal{X}}
\def\Y{\mathcal{Y}}
\def\Z{\mathcal{Z}}
\def\W{\mathcal{W}}
\def\U{\mathcal{U}}

\def\D{\mathcal{D}}
\def\balpha{\mbox{\boldmath $\alpha$}}

\def\th{^{\mbox{\scriptsize th}}}

\def\b\pi{\mbox{\boldmath $\pi$}}

\def\tr{\text{tr}}
\def\Vec{\text{vec}}

\newtheorem{theorem}{Theorem}[section]
\newtheorem{lemma}[theorem]{Lemma}
\newtheorem{condition}[theorem]{Condition}
\newtheorem{corollary}[theorem]{Corollary}

\newtheorem{definition}[theorem]{Definition}

\usepackage{xr}
\endlocaldefs

\begin{document}

\begin{frontmatter}

\title{The EAS approach for graphical selection consistency in vector autoregression models}
\runtitle{EAS for VAR graph selection}

\author{\fnms{Jonathan P} \snm{Williams}\thanksref{m1}\ead[label=e1]{jwilli27@ncsu.edu}},
\author{\fnms{Yuying} \snm{Xie}\thanksref{m2}\ead[label=e2]{xyy@msu.edu}}, 
\and
\author{\fnms{Jan} \snm{Hannig}\thanksref{m1}\ead[label=e3]{jan.hannig@unc.edu}}
\address{\printead{e1}}
\affiliation{University of North Carolina at Chapel Hill\thanksmark{m1} Michigan State University\thanksmark{m2}}

\runauthor{J. Williams}

\begin{abstract}
As evidenced by various recent and significant papers within the frequentist literature, along with numerous applications in macroeconomics, genomics, and neuroscience, there continues to be substantial interest to understand the theoretical estimation properties of high-dimensional vector autoregression (VAR) models.  To date, however, while Bayesian VAR (BVAR) models have been developed and studied empirically (primarily in the econometrics literature) there exist very few theoretical investigations of the repeated sampling properties for BVAR models in the literature.  In this direction, we construct methodology via the $\varepsilon$-$admissible$ subsets (EAS) approach for posterior-like inference based on a generalized fiducial distribution of relative model probabilities over all sets of active/inactive components (graphs) of the VAR transition matrix.  We provide a mathematical proof of $pairwise$ and $strong$ graphical selection consistency for the EAS approach for stable VAR(1) models which is robust to model misspecification, and demonstrate numerically that it is an effective strategy in high-dimensional settings. 

\end{abstract}

\begin{keyword}
empirical Bayes, generalized fiducial inference, graph selection, high-dimensional model selection, selection consistency, vector autoregression
\end{keyword}

\end{frontmatter}

\section{Introduction}\label{VAR_intro}

Despite the lack of theoretical investigations of the repeated sampling properties for BVAR models, Bayesian methodology can surely offer important contributions to the high-dimensional VAR model literature, beyond what could be developed in a frequentist framework.  One notable such contribution is the construction of posterior distributions over the set of all relative model probabilities.  This framework of posterior inference has been widely exploited over the last decade in the high-dimensional linear regression literature, and we anticipate it will see comparable success for high-dimensional VAR models in the near future.  

Our constructed EAS methodology allows for such posterior-like inference of relative model probabilities for all graphs, and additionally we provide an algorithm which is self-tuning (i.e., no cross-validation is needed for calibration to data sets).  Such Bayesian model selection approaches are very useful for learning important relationships among the various components (univariate time-series) in the VAR model.  The EAS methodology is an entirely new perspective on model selection which was originally developed to effectively account for linear dependencies among subsets of covariates in the high-dimensional linear regression setting in \cite{Williams2019}.

To the best of our knowledge, our established $pairwise$ and $strong$ model selection consistency results are the first of their kind in the BVAR literature.  This type of result is sure to be followed by similar results in the high-dimensional BVAR literature, analogous to the emergence of model selection strong consistency results in the high-dimensional Bayesian linear regression literature such as \cite{Johnson2012, Narisetty2014, Williams2019}.

Further, we demonstrate how to construct an alternative framework for posterior-like inference in the VAR(1) model setting which eliminates prior choice and specification.  We avoid the necessity of prior distributions altogether by implementing a generalized fiducial inference (GFI) approach (see \citealt{Hannig2016}).  And while our model selection consistency results derive from a Gaussian assumption on the VAR(1) model errors, they are actually the first ever results about a fiducial distribution under model misspecification.  This is due to the fact that all of the supporting theorems and lemmas we contribute are non-asymptotic, and rely on a collection of explicit fourth moment bounds given in Section \ref{VAR_standalone_results}.  Consequently, as long as the VAR(1) model errors are independent within and across time and there exist bounded fourth moments, our generalized fiducial consistency results (which assume Gaussian data) still hold even if the true data is not Gaussian. 

We validate our methods empirically in low and high-dimensional settings on both synthetic and real data, and provide Python code for implementing our algorithm.  This code, and the workflow for reproducing all numerical results can be found at \verb1https://jonathanpw.github.io/research1.

Fiducial inference has a long history, but in the last decade there has been a renewed interest in the topic with a large number of authors contributing fundamental insights \citep{EdlefsenLiuDempster2009,BergerBernardoSun2009,  XieSingh2013, TaraldsenLindquist2013, VeroneseMelilli2014, MartinLiu2015book, schweder2016confidence, Fraser2019}. A gentle introduction to technical aspects of GFI is provided in Section \ref{VAR_methodology}.

Recent theoretical work on VAR models is largely comprised of considerations of regularized estimation procedures, most notably \cite{Basu2015}.  The Bayesian literature has not yet caught up.  There do exist numerous papers on BVAR methodology, especially in the econometric literature, but on predominantly empirical investigations, see for example \cite{Banbura2010, Korobilis2013, Giannone2015, Ahelegbey2016}.  The primary tool of the BVAR literature has been implementations of the Minnesota (shrinkage) prior and its variants \citep{Litterman1986}.  

It has been found that BVAR with shrinkage priors is effective for large VAR models of economic time-series, but little has been provided in the way of theoretical guarantees (a notable exception is \citealt{Ghosh2018}) or even uncertainty quantification of competing model choices (a notable exception is \citealt{Korobilis2013}).  
To the best of our knowledge, \cite{Ghosh2018} is the first in the literature to establish posterior parameter estimation consistency in the ``large p large n''  BVAR setting with $p = o(n)$, where $p$ is the dimension of the VAR model and $n$ is the number of observed time instances.  While their consistency results are about the posterior behavior of the transition matrix coefficients under various prior specifications, our consistency results are about the posterior-like behavior of all relative model probabilities (akin to Bayes factors) under the prior-free GFI framework.

We loosely adopt notation for multivariate time-series from \cite{Lutkepohl2005}.  The time-series $X^{(1)},\dots,X^{(n)} \in \R^{p}$ is taken to denote data from a VAR(1) model with no serial correlation, and so is generated as
\begin{equation}\label{VAR_model}
\Y = A\X + \Sigma^{\frac{1}{2}}\U,
\end{equation}
where 
$
\Y := \big(
\begin{smallmatrix}
X^{(1)} & \cdots & X^{(n)} \\
\end{smallmatrix} \big)
$ and $
\X := \big(
\begin{smallmatrix}
X^{(0)} & \cdots & X^{(n-1)} \\
\end{smallmatrix} \big)
$ are $p\times n$ matrices,
$\U := \big(
\begin{smallmatrix}
U^{(1)} & \cdots & U^{(n)} \\
\end{smallmatrix} \big)
$ is a $p\times n$ matrix with $U^{(t)} \overset{\text{iid}}{\sim} \text{N}_{p}(0,I_{p})$ for $t \in \{1,\dots,n\}$, $A$ is a $p\times p$ matrix of coefficients, and $\Sigma := \text{diag}\{\sigma_{1}^{2},\dots,\sigma_{p}^{2}\}$.  Assume $X^{(0)}$ is the $p$-dimensional zero vector.  Further, let $G \subseteq \{1,\dots,p^{2}\}$ be a set of indices denoting a graph of active components of $A$, and take $A_{g}$ to be the $p\times p$ matrix $A$ with active components corresponding to the graph $G$ (all other components are zero).  

We extend the high-dimensional linear regression EAS methodology developed in \cite{Williams2019} to this VAR(1) setting.  The idea behind the EAS procedure is to efficiently make inference on the set of $2^{p^{2}}$ graphs, $G$, by discriminating on graphs which contain redundant active components.  Our notion of redundancy is defined rigorously by the `$h$-function' given later in (\ref{VAR_h_function_VAR1}).  

However, the basic intuition is to assign negligible posterior-like probability to all $A_{g}$ that can be closely approximated, predictively, by a graph containing fewer active components.  This can occur for a variety of reasons, namely, correlated time-series in the VAR system of equations, and too small signal-to-noise coefficient magnitudes.  For example, suppose 
$G = \{1,2,3,4\}$ with 
$A_{g} = \big(
\begin{smallmatrix}
a_{11} & a_{12} \\
a_{21} & a_{22} \\
\end{smallmatrix}\big)$.
Then the coefficient matrix $A_{g}$ is not $\varepsilon$-$admissible$ if, for instance, for some well-calibrated precision, $\varepsilon > 0$, 
\[
\left\|
\big(\begin{smallmatrix}
a_{11} & a_{12} \\
a_{21} & a_{22} \\
\end{smallmatrix}\big) \X - 
\big(\begin{smallmatrix}
a_{11} & 0 \\
a_{21} & a_{22} \\
\end{smallmatrix}\big) \X \right\| < \varepsilon,
\]
where $\|\cdot\|$ is some measure of distance.  In this case, predictions from the graph $\{1,2,4\}$ approximate that of $A_{g}$ within $\varepsilon$ precision, and so $A_{g}$ is said to contain redundant information.  

Note that in finite samples, and particularly high-dimensional, settings with highly-correlated data the EAS framework has the intuition that the oracle graph itself may not be $\varepsilon$-$admissible$.  In these settings, the EAS methodology re-defines the notion of the `true' graph to be some non-redundant subgraph of the oracle graph, at least non-asymptotically.  This idea is important because it suggests that to develop inherently scalable methodology the key may be to re-define the notion of what one should hope to recover from a `true' data generating model in high-dimensional settings.  Additional intuition for the EAS methodology is provided in \cite{Williams2019} in the context of linear regression.

The remainder of the paper is organized as follows.  Section \ref{VAR_methodology} defines the notion of $\varepsilon$-$admissibility$ as well as constructs the generalized fiducial distribution for the EAS approach, and describes the Markov chain Monte Carlo (MCMC)-based computations.  The main theoretical results are presented in Section \ref{VAR_theory}, and numerical results are provided in Sections \ref{VAR_simulation} and \ref{VAR_real_data}.  
The majority of the proofs are moved to the supplementary materials.

\section{Methodology}\label{VAR_methodology}

To adapt ideas more smoothly from the linear regression setting of \cite{Williams2019}, re-express the VAR(1) model in (\ref{VAR_model}) in the form
\begin{equation}\label{VAR_DGE}
Y = \Z_{G_{o}} \balpha_{G_{o}}^{0} + (\W^{0})^{\frac{1}{2}}\Vec(\U),
\end{equation}
where $Y := \Vec(\Y)$, $\Z := \X'\otimes I_{p}$, $\W^{0} := I_{n}\otimes \Sigma^{0}$, $\balpha := \Vec(A)$, and $G_{o}$ (as well as $g_{o}$ seen later) denotes the oracle graph.  Here and throughout, the superscript-zero notation denotes the true fixed values of the corresponding quantities.  The subscript notation, $\Z_{G_{o}}$ (or $\balpha_{G_{o}}$), refers to the sub-matrix (or sub-vector) with columns (or components) corresponding to the active components given by the index set $G_{o}$.  The $\Vec(\cdot)$ operator transforms an $n\times p$ matrix into an $np\times1$ vector by stacking columns in descending order, from left to right.  For example, $\Vec(\Y) = 
(\begin{smallmatrix}
X^{(1)'} & \cdots & X^{(n)'} \\
\end{smallmatrix})'$.  This linear model representation is also more convenient for expressing the likelihood function,
\begin{equation}\label{VAR_likelihood}
f\big(Y | \balpha_{G_{o}}, \{\sigma_{j}\}\big) = \frac{1}{  (2\pi)^{\frac{np}{2}} \big(\sigma_{1}^{2}\cdots\sigma_{p}^{2}\big)^{\frac{n}{2}} } e^{-\frac{1}{2} (Y - \Z_{G_{o}} \balpha_{G_{o}})'\W^{-1}(Y - \Z_{G_{o}} \balpha_{G_{o}}) },
\end{equation}
which will be needed later on.  For conciseness, the notation $\{\sigma_{j}\}$ is used as shorthand for $\{\sigma_{1},\dots,\sigma_{p}\}$.

Additional notation used for the remainder of the paper includes the following.  For a scalar-valued argument $|\cdot|$ represents the absolute value, but for a set-valued argument it represents the cardinality.  The norms $\|\cdot\|$ and $\|\cdot\|_{0}$ denote the vector $L_{2}$ and $L_{0}$ norms, respectively, while for a matrix $A$, $\|A\|_{2} := \sqrt{\lambda_{\max}(A'A)}$ and $\|A\|_{F} := \sqrt{\tr(A'A)}$ represent the matrix spectral and Frobenius norms, respectively.  Additionally, the quantities $\lambda_{\min}(A)$ and $\lambda_{\max}(A)$ denote the minimum and maximum eigenvalues of a given matrix, $A$, respectively.  The notation $P(\cdot)$ and $E(\cdot)$ refer, respectively, to the probability measure and expectation with respect to the joint generalized fiducial distribution of $A_{g}$ and $\Sigma$.  Conversely, the notation $P_{x}(\cdot)$ and $E_{x}(\cdot)$ refer, respectively, to the probability measure and expectation associated with the uncertainty from the VAR(1) process, rather than the probability measure for the generalized fiducial distribution of the unknown parameters.

The centerpiece of the EAS model selection approach is a definition of model redundancy, as made rigorous by our notion of $\varepsilon$-$admissibility$ and the $h$-function, presented next.  As described in Section \ref{VAR_intro}, the main intuition is that $\balpha_{G}$ is considered non-redundant, or $\varepsilon$-$admissible$, if and only if there does not exist a close fitting graph with strictly fewer active components.  However, there are also two additional constraints embedded in the $h$-function for $\varepsilon$-$admissibility$. 
\begin{definition}
Assume $\varepsilon, d > 0$ and $c \in (0,1)$.  A given coefficient matrix $A_{g}$, equivalently $\balpha_{G}$, for some graph $G$ is said to be $\varepsilon$-$admissible$ if and only if $h\big(\balpha_{G}, \{\sigma_{j}\}\big) = 1$, where
\begin{equation}\label{VAR_h_function_VAR1} 
h\big(\balpha_{G}, \{\sigma_{j}\}\big) := 1\Bigg\{\frac{1}{2}\|\Z_{G}'\W^{-1}\Z_{G}(\balpha_{G} - b_{\min})\|^{2} \ge \varepsilon, \min_{1\le j\le p}\{m_{j}^{g}\} \ge d, \|A_{g}\|_{2} \le c \Bigg\}
\end{equation}
where $b_{\min}$ solves $\ds \min_{b\in\R^{|G|}} \frac{1}{2}\|\Z_{G}'\W^{-1}\Z_{G}(\balpha_{G} - b)\|^{2} \text{ subject to } \|b\|_{0} \le |G|-1$,
\begin{equation}\label{VAR_Sigma_est_diag}
\{m_{1}^{g}, \dots, m_{p}^{g}\} = \text{diag}\big\{(\Y - \widehat{A}_{g}\X)(\Y - \widehat{A}_{g}\X)'\big\},
\end{equation}
and $\widehat{A}_{g} := \Y \Z_{G}' (\Z_{G}\Z_{G}')^{-1}$ is the least squares estimator for graph $G$.
\end{definition}
To begin to understand the behavior of the $h$-function, first note that
\[
\|\Z_{G}'\W^{-1}\Z_{G}(\balpha_{G} - b_{\min})\|^{2} = \|\Z_{G}'\W^{-1}(\Z_{G}\balpha_{G} - \Z_{G}b_{\min})\|^{2},
\]
is analogous to a noiseless version of the Dantzig selector \citep{Candes2007} where $\Z_{G}$ is the design matrix for the linear model representation (\ref{VAR_DGE}).  One reason to use $\Z_{G}'\W^{-1}\Z_{G}$ versus simply $\Z_{G}$ is that the former is scale-invariant to the $\{\sigma_{j}\}$ and invariant to orthogonal transformations of the data.  Second, note that if $\Z_{G}$ contains linearly dependent columns, then for any coefficients $\balpha_{G}$, the linear prediction $\Z_{G}\balpha_{G}$ can be exactly recovered by $\Z_{G}b_{\min}$ (since $\|b_{\min}\|_{0} \le |G|-1$).  This immediately implies that since $\Z_{G}$ is an $np\times|G|$ matrix, for all $G$ with $|G| > np$, $h\big(\balpha_{G}, \{\sigma_{j}\}\big) = 0$ by definition.  For high-dimensional settings where $p > n$, then by construction, considering only $\varepsilon$-$admissible$ graphs reduces the model selection problem from $2^{p^{2}}$ candidate graphs to only $2^{np}$.  This fact makes the EAS methodology inherently scalable.

The quantities $c$, $d$, and $\varepsilon$ will now be described in alphabetical order.  The component, $\|A_{g}\|_{2} \le c$, in the $h$-function concentrates the distribution of $A_{g}$ to only allow for stable VAR(1) models with $c \in (0,1)$.  In practice, since $\|A^{0}\|_{2}$ is typically not known the constraint $\|A_{g}\|_{2} \le c$ is replaced by $\|A_{g}\|_{2} < 1$.  The second component in the $h$-function is the expression $\min_{1\le j\le p}\{m_{j}^{g}\} \ge d$, where $m_{j}^{g}$ for $j \in \{1,\dots,p\}$ is understood as the residual sum-of-squares (RSS) for the $j\th$ component of the VAR system.  The basic idea is that the data-dependent quantity $d = d(\Y,\X,G_{o})$ should be calibrated to $\min_{1\le j\le p}\{m_{j}^{g_{o}}\}$ which corresponds to the oracle graph, and so any graphs which have a better fit than the oracle will be excluded from consideration via the $h$-function.  Accordingly, this device is designed to eliminate graphs which over-fit the data, and is important for establishing our asymptotic consistency results.  However, in practice $d$ can be set to a small value and left alone; more will be said about this in Section \ref{VAR_simulation} with the numerical results.  

For $\Z_{G}$ which have full column rank, the degree to which the features associated with graph $G$ are redundant depends on the correlations between the $p$ components of the VAR model, the distribution of the coefficients $\balpha_{G}$ (i.e., the transition matrix $A_{g}$), scale matrix components $\{\sigma_{j}\}$, and the specified level of precision, $\varepsilon$.  Our proposed default choice of $\varepsilon$, formulated from theoretical investigations (based on the Gaussian contemporaneous errors assumption), is for some $\rho \in (0,\frac{1}{2})$,
\begin{equation}\label{VAR_default_eps}
\varepsilon = \Lambda_{g} \cdot \max\Big\{ 1, n^{1-\rho}p^{2}\Big(.5\log(\log(n))|G| - |G_{o}|\Big)\Big\}.
\end{equation}
There are predominantly two components to $\varepsilon$; the quantity $\Lambda_{g} := \|\W^{-\frac{1}{2}}\Z_{G}\|_{F}^{2}$ is particularly calibrated to the observed data since it originates from a tight concentration inequality for the transition matrix $A_{g}$, and the term $n^{1-\rho}p^{2}\log(\log(n))|G|$ is necessary asymptotically for managing the accumulating data and rapidly growing number of candidate graphs as $n, p \to \infty$.  The basic idea is that $\Lambda_{g}$ will always contribute, and the remaining terms will contribute for sufficiently large $n$ or for $|G|$ which exceeds the number of active components in the oracle model.  However, as is demonstrated in Section \ref{VAR_simulation}, for observed data $\Lambda_{g}$ is so well-calibrated that it suffices to set $\varepsilon = \Lambda_{g}$, and thus also eliminating the need for a tuning parameter.  More details about $\Lambda_{g}$ are given in Section \ref{VAR_theory_results}, particularly its expectation in (\ref{VAR_exp_value_Lambda}).  

With the EAS methodology now developed a framework of statistical inference is required for implementing it.  A suitable such framework is GFI because it will allow us to construct posterior-like inference over the $2^{p^{2}}$ candidate graphs without having to specify any prior distributions.  The intuition for GFI is to begin with a data generating equation such as (\ref{VAR_DGE}) and invert the equation on the data to solve for the unknown parameters.  The resulting quantity is defined as the generalized fiducial distribution of the unknown parameters.  Precise details for the construction of this approach are provided in \cite{Hannig2016}.  The generalized fiducial probability density function for the parameters in the VAR(1) model (\ref{VAR_DGE}) has the form
\begin{equation}\label{VAR_GFdensity}
r(\balpha_{G},\{\sigma_{j}\} \mid Y) = \frac{    f\big(Y \mid \balpha_{G}, \{\sigma_{j}\}\big) \cdot J\Big(Y, \big(\balpha_{G}, \{\sigma_{j}\}\big) \Big) \cdot h\big(\balpha_{G}, \{\sigma_{j}\}\big)    }{    \int\int f\big(Y \mid \balpha_{G}, \{\sigma_{j}\}\big) \cdot J\Big(Y, \big(\balpha_{G}, \{\sigma_{j}\}\big) \Big) \cdot h\big(\balpha_{G}, \{\sigma_{j}\}\big) \ d\balpha_{G} \ d\{\sigma_{j}\}   },
\end{equation}
where the multiplication by the $h$-function appears as an infusion of the EAS methodology into the GFI framework, and the Jacobian term,
\[
J\Big(Y, \big(\balpha_{G}, \{\sigma_{j}\}\big) \Big) := D\bigg( \nabla_{(\balpha_{G}, \{\sigma_{j}\})} V\big(u,(\balpha_{G}, \{\sigma_{j}\})\big)\Big|_{u=V^{-1}\big(Y,(\balpha_{G}, \{\sigma_{j}\})\big)} \bigg)
\]
with $D(A) = (\det A'A)^{\frac{1}{2}}$ and $V$ denoting the data generating equation (\ref{VAR_DGE}).  The Jacobian term results from inverting the data generating equation on the unknown parameters.  Note that the $\{\sigma_{j}\}$ are also dependent on the the particular graph $G$, but this dependence is suppressed in the notation for conciseness.

The likelihood function in (\ref{VAR_GFdensity}) is given by (\ref{VAR_likelihood}), the $h$-function is given by (\ref{VAR_h_function_VAR1}), and the derivation of the Jacobian term is presented in the supplementary material.  From the generalized fiducial density of $\balpha_{G}$ and $\{\sigma_{j}\}$, the generalized fiducial mass function for a graph $G$ is proportional to the normalizing constant in (\ref{VAR_GFdensity}).  In Bayesian theory, this constant of proportionality is understood as the marginal density of the data.  Evaluating the integral in the denominator of (\ref{VAR_GFdensity}) gives,
\begin{equation}\label{VAR_pmf}
r(G \mid Y) \propto \frac{   E\Big( h\big(\balpha_{G}, \{\sigma_{j}\}\big)|\widetilde{\D}'_{g}\widetilde{\D}_{g}|^{\frac{1}{2}} \Big)    \prod_{j=1}^{p} \big(\frac{m_{j}^{g}}{2}\big)^{-\frac{n-|r_{j}^{g}|}{2}} \Gamma\Big(\frac{n-|r_{j}^{g}|}{2}\Big)   }{ (\frac{n}{2\pi})^{\frac{|G|}{2}} \prod_{j=1}^{p} \Big|\sum_{t=1}^{n} X_{r_{j}^{g}}^{(t-1)}X_{r_{j}^{g}}^{(t-1)'}\Big|^{\frac{1}{2}}  },
\end{equation}
where $r_{j}^{g}$ is the set of active row indices of $A_{g}$ for column $j \in \{1,\dots,p\}$, and $\widetilde{\D}_{g}$ is a data-dependent and parameter-free $(np)\times(|G|+p)$ matrix defined in the supplementary material as part of the Jacobian term.  Note that the inner expectation is with respect to the N$_{|G|}\big(\widehat{\balpha}_{g}, (\Z_{G}'\W^{-1}\Z_{G})^{-1}\big)$ distribution, conditional on $\{\sigma_{j}^{2}\}$, and for each $\sigma_{j}^{2}$, is taken with respect to the inv-gamma$\big( \frac{1}{2}(n-|r_{j}^{g}|), \frac{1}{2}m_{j}^{g}\big)$ distribution.  To ensure that $r(G \mid Y)$ defines a proper probability mass function, the normalizing constant in (\ref{VAR_pmf}) is scaled so that $\sum_{i=1}^{p^{2}}\sum_{G : |G|=i} r(G \mid Y) = 1$.

Lastly, the relative model probabilities (\ref{VAR_pmf}) can be computed via psuedo-marginal MCMC algorithms.  Traditional MCMC is not feasible because the expected value appearing in (\ref{VAR_pmf}) is not available in closed form.  We implement the grouped independence Metropolis-Hastings (GIMH) algorithm described in \citep{Andrieu2009}, which replaces the expected value with the empirical mean of importance samples at each step of the MCMC algorithm.  In the case of (\ref{VAR_pmf}), efficient importance samples are easily drawn from the N$_{|G|}\big(\widehat{\balpha}_{g}, (\Z_{G}'\W^{-1}\Z_{G})^{-1}\big)$ and inv-gamma$\big( \frac{1}{2}(n-|r_{j}^{g}|), \frac{1}{2}m_{j}^{g}\big)$ distributions for $\balpha_{G}$ and $\sigma_{j}$, respectively.  The GIMH algorithm we construct is a Markov chain on the set of graphs $G \subseteq \{1,\dots,p^{2}\}$, and proposals are made by either adding, removing, or replacing a component index in the current iterate of $G$ in the chain.

A point of caution about the GIMH algorithm is that the mixing conditions are usually particularly sensitive to the number of importance samples taken to estimate an expectation at each step of the algorithm.  However, the algorithm mixed well enough to yield very encouraging numerical results for the high-dimensional linear regression setting in \citep{Williams2019}, and Sections \ref{VAR_simulation} and \ref{VAR_real_data}, here, serve to demonstrate that the algorithm is not only computationally feasible but also favorable for graph selection in the VAR(1) model setting.  Further discussion of the algorithm is provided in \citep{Williams2019}, and a detailed pseudo-code description of the algorithm is provided at \verb1https://jonathanpw.github.io/research1.

\section{Theoretical results}\label{VAR_theory}

The problem of graphical selection is difficult because the number of candidate graphs to choose among grows super-exponentially in the dimension of the VAR(1) model, $2^{p^{2}}$.  Accordingly, the utility of the EAS procedure is its inherent ability to effectively manage a very large number of candidate graphs by assigning negligible posterior-like probability to redundant graphs.  The meaning of this assertion is made precise in Theorem \ref{VAR_MainResult} which states that the generalized fiducial distribution obtained from the EAS methodology exhibits pairwise graph selection consistency as both $n$ and $p$ are taken to infinity, and as a corollary, strong selection consistency for fixed $p$.  The necessary mathematical conditions are discussed next.

\subsection{Conditions}

The first two conditions presented are related to the identifiability of the true data generating graph, $G_{o}$.  We consider only a stable VAR(1) model for our theoretical investigation, and adopt the common notion of stability that for the true transition matrix $\|A^{0}\|_{2} \le c$ for some $c \in (0,1)$.  It is assumed throughout that a valid $c$ has been fixed a-priori.

Condition \ref{VAR_condition1} arises in the proof of Lemma \ref{VAR_JacobianLowBLemma} which is a necessary result for Theorem \ref{VAR_TrueGraphLowBTheorem}.  It guarantees that the Jacobian term for the oracle graph in (\ref{VAR_pmf}) will be lower bounded away from zero in probability.  The quantity $\delta$ represents an approximation to $\lambda_{\min}\big(\Omega - E_{x}(\Omega)\big)$ (via Lemma \ref{VAR_omega_bound}) which manages the uncertainty resulting from the minimum eigenvalue of the Jacobian matrix $\widetilde{\D}'_{g_{o}}\widetilde{\D}_{g_{o}}$, where
$
\Omega := \frac{1}{n} 
\big(\begin{smallmatrix}
\X\X' & \X\U'  \\
\U\X' & \U\U' \\
\end{smallmatrix}\big)
$ 
and 
$
E_{x}(\Omega) = 
\Big(\begin{smallmatrix}
\Gamma_{n}(0) &  \\
 & I_{p} \\
\end{smallmatrix}\Big)
$.  It is also assumed that a valid $\delta > 0$ has been fixed a-priori.
\begin{condition}\label{VAR_condition1}
The true transition matrix satisfies $\|A^{0}\|_{2} \le c < 1$, $\lambda_{\max}\big(\Gamma_{n}(0)\big)$ is bounded from above by a fixed constant, and
\[
\sqrt{n}\left[
\lambda_{\min}
\begin{pmatrix}
\Gamma_{n}(0) &   \\
 & I_{p} \\
\end{pmatrix} - \delta\right] > 4(1 + c^{2}),
\]
where $\delta > 0$, and
\[
\Gamma_{n}(0) := \frac{1}{n}E_{x}(\X\X') = \frac{1}{n}\sum_{t=1}^{n}\sum_{k=0}^{t-2}(A^{0})^{k}\Sigma^{0}(A^{0})^{k'}.
\]
Observe that this condition also implies that $\lambda_{\min}\big(\Gamma_{n}(0)\big) > \delta$.
\end{condition}
Note that Lemma \ref{VAR_omega_bound} guarantees $\lambda_{\min}\big(\Omega - E_{x}(\Omega)\big) \overset{P_{x}}{\longrightarrow} 0$ as $n \to \infty$, assuming the $p$ versus $n$ relationship given by Condition \ref{VAR_condition4}.  Thus, the condition can reasonably be verified on real data by assuming $\delta > 0$ is arbitrarily small and comparing the value of 
$
\sqrt{n}\lambda_{\min}
\Big(\begin{smallmatrix}
\frac{1}{n}\X\X' &  \\
 & I_{p} \\
\end{smallmatrix}\Big)$
to $4(1 + c^{2})$, where $\frac{1}{n}\X\X'$ is the obvious sample analogue to the population quantity considered in Condition \ref{VAR_condition1}.  Since $c$ is unknown in practice, for the purposes of checking this condition on real data evaluate $4(1 + c^{2}) = 8$ for the worst case with $c$ replaced by 1.  We demonstrate on synthetic data in Section \ref{VAR_simulation} that this verifiable condition is indeed meaningful for practical applications.

Condition \ref{VAR_condition2}, which originates from the proof of Theorem \ref{VAR_TrueGraphLowBTheorem}, is also well calibrated to real data.  This condition states the maximum rate at which $\varepsilon$ can be allowed to grow as a function of $n, p$, and $\Lambda_{g_{o}}$, whilst the oracle model $G_{o}$ remains identifiable (i.e., no faster than $n^{1-\rho}p^{2}\Lambda_{g_{o}}$).  The fixed quantity $\rho \in (0,\frac{1}{2})$ represents the `gap' between how fast $\varepsilon$ must grow (stated in Condition \ref{VAR_condition4}) to effectively manage the set of all $2^{p^{2}}$ candidate graphs under consideration, and how slow it must grow to not eliminated the oracle graph from consideration.  Namely, $\varepsilon \propto n^{1-\rho}p^{2}\Lambda_{g}$ simultaneously satisfies Conditions \ref{VAR_condition2} and \ref{VAR_condition4} for any $\rho \in (0,\frac{1}{2})$.  It is assumed throughout that a valid $\rho$ has been fixed a-priori.  The quantities on the left side of the inequality in Condition \ref{VAR_condition2} are expected values of the corresponding quantities on the left side of the first constraint in the $h$-function (\ref{VAR_h_function_VAR1}). 
\begin{condition}\label{VAR_condition2}
The oracle graph, $G_{o}$, satisfies $\min_{1\le j\le p}\{m_{j}^{g_{o}}\} \ge d$,
\[
\frac{1}{18}\big\|(\Gamma_{n}(0)\otimes (\Sigma^{0})^{-1})_{G_{o},G_{o}} (\balpha_{G_{o}}^{0} - \widetilde{b})\big\|^{2} \ge \frac{\varepsilon}{n^{1-\rho}p^{2}\Lambda_{g_{o}}},
\]
where $\rho \in (0,\frac{1}{2})$, $\widetilde{b}$ solves $\min_{b\in\R^{|G_{o}|}}\big\|(\Gamma_{n}(0)\otimes (\Sigma^{0})^{-1})_{G_{o},G_{o}} (\balpha_{G_{o}}^{0} - b)\big\|^{2}$ subject to $\|b\|_{0} \le |G_{o}|-1$, and $\varepsilon = \Lambda_{g_{o}}\cdot \widetilde{\varepsilon}$ for some $\widetilde{\varepsilon}$ not depending on $\Sigma$ or $A_{g_{o}}$.
\end{condition}
Unless the oracle model is known, Condition \ref{VAR_condition2} is not verifiable on real data, but in Section \ref{VAR_simulation} we are able to demonstrate the varying performance of the EAS procedure on simulated data when this condition is and is not satisfied.  Note, that the coefficient of $\frac{1}{18}$ is a constant more pertinent to asymptotic considerations (and our proof technique), and should be understood as closer to the value of $\frac{1}{2}$ (which appears in the $h$-function).

The next condition is a component in the proof of Theorem \ref{VAR_BigGraphUpBTheorem} for guaranteeing that the $h$-function will drive the EAS procedure to assign negligible posterior-like probability to non-$\varepsilon$-$admissible$ graphs, $G$, via the mass function $r(G \mid Y)$ in (\ref{VAR_pmf}).
\begin{condition}\label{VAR_condition3}
For any $G$ with $G \not\subseteq G_{o}$,
\[
\frac{9}{2}\big\|\big(E_{x}(\Z'_{G}\Z_{G})\big)^{-1}E_{x}(\Z_{G}'Y) - \widetilde{b}\big\|^{2} < \frac{\varepsilon}{n^{1+\frac{\rho}{2}} p^{3}\Lambda_{g}},
\]
where $\widetilde{b}$ solves $\min_{b\in\R^{|G|}}\big\|\big(E_{x}(\Z'_{G}\Z_{G})\big)^{-1}E_{x}(\Z_{G}'Y) - b\big\|^{2}$ subject to $\|b\|_{0} \le |G|-1$, and $\varepsilon = \Lambda_{g}\cdot \widetilde{\varepsilon}$ for some $\widetilde{\varepsilon}$ not depending on $\Sigma$ or $A_{g}$.
\end{condition}
The intuition for Condition \ref{VAR_condition3} is that for graphs containing redundant active components the central tendency of the least squares estimator $\widehat{\balpha}_{g}$ can be closely approximated by a vector of fewer active components.  Notice that $\big(E_{x}(\Z'_{G}\Z_{G})\big)^{-1}E_{x}(\Z_{G}'Y)$ is an approximation to $E_{x}(\widehat{\balpha}_{g})$.  Since the least squares estimator is asymptotically well behaved for Gaussian VAR models, this condition is not particularly interesting and is easily satisfied in numerical experiments.  Furthermore, it will hold trivially, for instance, if the columns $\Z_{G}$ are linearly dependent.

The final condition in this section is Condition \ref{VAR_condition4}, which simply states the asymptotic rate at which $\varepsilon$ and $d$ from the definition of $h$ in \eqref{VAR_h_function_VAR1} must increase as $n, p \to \infty$ for our main result, Theorem \ref{VAR_MainResult}, to be established.  In fact, the previous three conditions were all for establishing non-asymptotic bounds of concentration.
\begin{condition}\label{VAR_condition4}
For some fixed $\rho \in (0,\frac{1}{2})$, $p^{\max\big\{\frac{14}{\rho},\frac{2}{1-2\rho}\big\}} = o(n)$.  For the positive constant $K_{1}$ specified in (\ref{VAR_K1}), as $n \to \infty$ or $n, p \to \infty$, $\varepsilon$ satisfies
\[
\frac{\varepsilon}{9\Lambda_{g}} - K_{1} \bigg(\frac{p\|Y\|^{2}}{\sqrt{n}} + p^{2}\log(n) + \frac{n}{q} \cdot p^{2}\sqrt{n}\bigg) \overset{P_{x}}{\longrightarrow} \infty,
\]
$d$ satisfies
\[
\frac{d\cdot n^{\frac{\rho}{2}} p^{2}}{4\lambda_{\max}(\X\X'/n)} - \frac{np}{2} - K_{1} \bigg(\frac{p\|Y\|^{2}}{\sqrt{n}} + p^{2}\log(n) + \frac{n}{q} \cdot p^{2}\sqrt{n}\bigg) \overset{P_{x}}{\longrightarrow} \infty,
\]
and $n = O_{p}(q)$, where $q := \min_{1\le j\le p}\{m_{j}\}$ with $m_{1},\dots,m_{p}$ corresponding to the full model (i.e., all components active), and $\varepsilon = \Lambda_{g}\cdot \widetilde{\varepsilon}$ for some $\widetilde{\varepsilon}$ not depending on $\Sigma$ or $A_{g}$.
\end{condition}
A important attribute of Condition \ref{VAR_condition4} is the requirement that while the dimension of the VAR(1) model, $p$, can be taken to infinity, it must be exceeded polynomially by the number of observed time instances, $n$.  This is in contrast to the model selection consistency result established for the high-dimensional linear regression setting in \citep{Williams2019}, where $p$ was allowed to grow sub-exponentially in $n$.  The primary difference here is that we derive model selection consistency results for the multivariate VAR model setting which are robust to model misspecification, namely the assumption of Gaussian VAR model errors.  Such a robust generalized fiducial result requires (to the best of our understanding) non-asymptotic second moment concentration bounds.  High-dimensional ($p > n$) consistency results require exponential tail bounds when establishing concentration of data-dependent quantities such as in Lemma \ref{VAR_bound2} in the next section, and exponential tail bounds here are intimately related to the assumption of Gaussianity.

Note that no assumption of sparsity is made in any of the conditions.  This section concludes with a definition of various quantities that will be referenced in the next section, and throughout the proofs.
\begin{definition}\label{VAR_def_quantities}
$N_{1}$ is any positive constant such that $n \ge N_{1}$ implies
\[
1 - \frac{1 - c^{2n}}{n(1-c^{2})} \le 1.
\]
$N_{2}$ is any positive constant such that $n \ge N_{2}$ implies
\[
1 + c^{2} - 2\frac{c^{2} - (c^{2})^{n+1}}{n(1-c^{2})} \le 1 + c^{2}.
\]
Additionally, $N_{3}$ is defined as in (\ref{VAR_TrueGraphLowBTheorem_final}).
\begin{equation}\label{VAR_V1}
\begin{split}
V_{1} & := 16(\sigma_{\max}^{0})^{4} \Bigg[ \frac{p^{6} n^{1-\frac{3\rho}{2}}}{\xi} \cdot \bigg( \frac{\|\Gamma_{n}(0)\|_{2}^{2}}{(\sigma_{\max}^{0})^{4} p} + \frac{ (3 + c^{4}) }{(1-c^{2})^{3}n} \bigg) \\
& \hspace{1.75in} + \frac{ \delta^{-2} p^{2}}{ (1-c^{2})^{3} n^{1-2\rho} } + \frac{ (3 + c^{4}) p^{6}n^{\frac{\rho}{2}}}{(1-c^{2})^{3}\xi} \Bigg], \\
\end{split}
\end{equation}
with $\xi = \frac{2\delta^{2}}{9\Lambda_{g}} \varepsilon$.  The alternate $\widetilde{V}_{1}$ denotes $V_{1}$ with $\varepsilon$ replaced by $c^{2}\cdot \frac{9 n^{1+\frac{\rho}{2}} p^{3} \Lambda_{g_{o}}}{2}$.
\begin{equation}\label{VAR_V2}
V_{2} := 4\delta^{-2} \frac{(\sigma_{\max}^{0})^{4} (1 + c^{2})}{(1 - c^{2})^{3}} \cdot \frac{2\min\{|G_{o}|,p\}^{2}}{n}.
\end{equation}
\begin{equation}\label{VAR_V3}
V_{3} :=  \frac{V_{2}}{4} + \delta^{-2}\Bigg[\frac{ 2p (\sigma_{\max}^{0})^{2} \min\{|G_{o}|, p\} }{ n (1 - c^{2}) } + \frac{p(p + 1)}{n} \Bigg].
\end{equation}
\end{definition}

\subsection{Results}\label{VAR_theory_results}

Our strategy for establishing graph selection consistency in Theorem \ref{VAR_MainResult} is largely composed of the contents of Lemmas \ref{VAR_bound1} and \ref{VAR_bound2} and Theorems \ref{VAR_BigGraphUpBTheorem} and \ref{VAR_TrueGraphLowBTheorem}.  Lemmas \ref{VAR_bound1} and \ref{VAR_bound2} describe, respectively, the generalized fiducial concentration of the VAR(1) transition matrix around its least squares estimate and the concentration of the least squares estimate around an approximation to its expectation.  The probability bounded in Lemma \ref{VAR_bound1} is with respect to the joint generalized fiducial distribution of $A_{g}$ and $\Sigma$.  In contrast, Lemma \ref{VAR_bound2} is a concentration inequality with respect to the data generating mechanism (\ref{VAR_DGE}) which derives its distribution from the errors $U^{(t)} \overset{\text{iid}}{\sim} \text{N}_{p}(0,I_{p})$ for $t \in \{1,\dots,n\}$.  In what follows we chose $\varepsilon = \Lambda_{g}\cdot \widetilde{\varepsilon}$ for some $\widetilde{\varepsilon}$ not depending on $\Sigma$ or $A_{g}$.

\begin{lemma}\label{VAR_bound1}
For any $G$ with $|G| \le np$,
\[
P\Big(\|\Z_{G}'\W^{-1}\Z_{G}(\balpha_{G}  - \widehat{\balpha}_{g})\|^{2} \ge \varepsilon\Big) \le \frac{  |G| \sqrt{2\Lambda_{g}}  }{ \sqrt{\pi\varepsilon}   }  e^{-\frac{\varepsilon}{2\Lambda_{g}}},
\]
where $\widehat{\balpha}_{g} := \big(\Z_{G}'\Z_{G}\big)^{-1}\Z_{G}'Y$, and $\Lambda_{g} := \|\W^{-\frac{1}{2}}\Z_{G}\|_{F}^{2}$.
\end{lemma}
Recall that $\Lambda_{g}$, which comes from the proof of this lemma, is a key component of our suggested default $\varepsilon$ in (\ref{VAR_default_eps}) and of Condition \ref{VAR_condition4}.  This results from the fact that $\varepsilon$ must control for $\Lambda_{g}$ in order to establish the well-behaved concentration of the generalized fiducial distribution of $\balpha_{G}$ which is exhibited by this lemma.  The $\W^{-\frac{1}{2}}$ plays the role of appropriately scaling the design matrix $\Z_{G}$.  Observe that for the full model $G = \{1,\dots,p^{2}\}$,
\[
\Lambda = \|\W^{-\frac{1}{2}}\Z\|_{F}^{2} = \tr(\Z'\W^{-1}\Z) = \tr\big((\X\X')\otimes \Sigma^{-1}\big) = \tr(\X\X')\cdot\tr(\Sigma^{-1}),
\]
which gives 
\begin{equation}\label{VAR_exp_value_Lambda}
E_{x}(\Lambda) = n\cdot \tr(\Gamma_{n}(0))\cdot \tr(\Sigma^{-1}).
\end{equation}
Thus, for a given graph $G$, $\Lambda_{g}$ is a combined measure of the covariance or dependence among the $p$ univariate time-series in the VAR model, the contemporaneous error precision matrix, and the number of observed instances of the time-series.  This is what makes $\Lambda_{g}$ effective as apart of $\varepsilon$ in the $h$-function for determining the $\varepsilon$-$admissibility$ of a given $\balpha_{G}$.  Lemma \ref{VAR_bound3} gives a probabilistic bound on $\Lambda_{g}$ as a function of $n$ and $p$, given the $h$-function constraint that $\min_{1\le j\le p}\{m_{j}^{g}\} \ge d$. 
\begin{lemma}\label{VAR_bound3}
For any $G$,
\[
P\bigg(\Lambda_{g} \ge n^{1+\frac{\rho}{2}}p^{3}, \min_{1\le j\le p}\{m_{j}^{g}\} \ge d\bigg) \le e^{-\big(\frac{d\cdot n^{\frac{\rho}{2}} p^{2}}{4\lambda_{\max}(\X\X'/n)} - \frac{np}{2}\big)} 2^{-\frac{|G|}{2}},
\]
where $\Lambda_{g} := \|\W^{-\frac{1}{2}}\Z_{G}\|_{F}^{2}$.
\end{lemma}

Next, consider the concentration of the least squares estimate.
\begin{lemma}\label{VAR_bound2}
Assume Condition \ref{VAR_condition1} holds.  Then for all $n \ge \max\{N_{1},N_{2}\}$, and for any $G$ with $|G| \le np$,
\[
P_{x}\bigg( \|\widehat{\balpha}_{g} - \big(E_{x}(\Z_{G}'\Z_{G})\big)^{-1}E_{x}(\Z_{G}'Y)\|^{2} \ge \frac{2\varepsilon}{9n^{1+\frac{\rho}{2}} p^{3} \Lambda_{g}} \bigg) \le V_{1},
\] 
where $V_{1}$ is as in (\ref{VAR_V1}).
\end{lemma}
Materially, the three preceding lemmas are needed in the proofs of Theorems \ref{VAR_BigGraphUpBTheorem} and \ref{VAR_TrueGraphLowBTheorem}, presented next.  These theorems are results about the behavior of the EAS methodology coupled with the generalized fiducial distribution (i.e., the Jacobian term); they are analogous to studying the behavior of given priors for a (Bayesian) posterior distribution.  Theorem \ref{VAR_BigGraphUpBTheorem} is a non-asymptotic concentration inequality which yields an upper bound on the rate at which the expected value (w.r.t. the joint generalized fiducial distribution of $A_{g}$ and $\Sigma$) of the $h$-function times the Jacobian term diverges for non-$\varepsilon$-$admissible$ graphs, $G$.
\begin{theorem}\label{VAR_BigGraphUpBTheorem}
Take any $G$ with $G \not\subseteq G_{o}$ and $|G| \le np$, and assume Conditions \ref{VAR_condition1} and \ref{VAR_condition3} hold.  Then for all $n \ge \max\{N_{1},N_{2}\}$,
\[
\begin{split}
E\Big( h\big(\balpha_{G}, \{\sigma_{j}\}\big)|\widetilde{\D}'_{g}\widetilde{\D}_{g}|^{\frac{1}{2}} \Big) & \le e^{\frac{1}{2}(1 - c)^{-2} \big(r_{\max}^{g} + (1 + c)^{2} \big)\frac{\|Y\|^{2}}{\sqrt{n}} - \frac{|G| + p}{2}} \\
& \times \bigg( \frac{  3|G| \sqrt{\Lambda_{g}}  }{ \sqrt{\pi\varepsilon}   }  e^{-\frac{\varepsilon}{9\Lambda_{g}}} + e^{-\big(\frac{d\cdot n^{\frac{\rho}{2}} p^{2}}{4\lambda_{\max}(\X\X'/n)} - \frac{np}{2}\big)} 2^{-\frac{|G|}{2}+1} \bigg) \\
\end{split}
\]
with probability exceeding $1 -V_{1}$, where $V_{1}$ is as in (\ref{VAR_V1}), $r_{\max}^{g} := \max_{1\le j\le p}|r_{j}^{g}|$.
\end{theorem}

Conversely, Theorem \ref{VAR_TrueGraphLowBTheorem} is a non-asymptotic lower bound on the $h$-function times the Jacobian term for the oracle graph, $G_{o}$.
\begin{theorem}\label{VAR_TrueGraphLowBTheorem}
Assume Conditions \ref{VAR_condition1}, \ref{VAR_condition2}, and \ref{VAR_condition4} hold.  Then for all $n \ge \max\{N_{1},N_{2},N_{3}\}$, with $N_{3}$ and the fixed $K_{3} \in (0,1)$ defined by (\ref{VAR_TrueGraphLowBTheorem_final}),
\[
P_{x}\Bigg( E\Big(h\big(\balpha_{G_{o}}, \{\sigma_{j}\}\big)|\widetilde{\D}'_{g_{o}}\widetilde{\D}_{g_{o}}|^{\frac{1}{2}}\Big) \ge (1 - K_{3}) e^{\frac{|G_{o}|+p}{4}} \Bigg) \ge 1 - V_{1} - \widetilde{V}_{1} - 2V_{2} - 2e^{-\frac{np}{4}} - V_{3},
\]
where $V_{1}$ and $\widetilde{V}_{1}$, $V_{2}$, and $V_{3}$ are as in (\ref{VAR_V1}), (\ref{VAR_V2}), and (\ref{VAR_V3}), respectfully.
\end{theorem}

Before stating the main result of this paper one final condition, Condition \ref{VAR_condition5}, is needed.  In its absence a less strong, yet still meaningful statement of posterior-like graphical consistency holds; we formulate this alternative statement as Corollary \ref{VAR_MainResult_corollary}.  The importance of Condition \ref{VAR_condition5} is that it covers the gap left open in Theorem \ref{VAR_BigGraphUpBTheorem} since the theorem only bounds the generalized fiducial probability of non-$\varepsilon$-$admissible$ graphs (i.e., $G \not\subseteq G_{o}$). 
\begin{condition}\label{VAR_condition5}
For the positive constant $K_{2}$ specified in (\ref{VAR_K2}),
\[
\max_{G : G\subset G_{o}} \Bigg\{  e^{K_{2} \big(\frac{p\|Y\|^{2}}{\sqrt{n}} + p^{2}\log(n)\big)} \prod_{j=1}^{p}\Bigg[ \frac{(m_{j}^{g_{o}})^{\frac{n-|r_{j}^{g_{o}}|}{2}}}{(m_{j}^{g})^{\frac{n-|r_{j}^{g}|}{2}}} \Bigg] \Bigg\} \overset{P_{x}}{\longrightarrow} 0
\]
as $n \to \infty$ or $n, p \to \infty$.
\end{condition}
Recall from (\ref{VAR_Sigma_est_diag}) that $m_{j}^{g}$ is the univariate RSS, corresponding to graph $G$, for the $j\th$ component of the VAR(1) model.  Hence, this condition is a statement that the product of the ratio of RSS components for the true graph over that of any strict sub-graph, taken to a power on the order of $n$, will vanish at a rate of $\text{exp}\big\{\frac{p\|Y\|^{2}}{\sqrt{n}}\big\} = O_{p}\big(\text{exp}\big\{p^{2}\sqrt{n}\big\}\big)$.  This is not unreasonable to expect since for each $j \in \{1,\dots,p\}$, $m_{j}^{g} = O_{p}(n)$, $m_{j}^{g_{o}} = O_{p}(n)$, $m_{j}^{g_{o}} \le m_{j}^{g}$ for $G \in \{G : G\subset G_{o}\}$, and an explicit condition about the oracle model being sufficiently better fitting than all sub-models is typical of model consistency results.

The main result of our paper, a statement of pairwise graphical selection consistency for the constructed EAS methodology, is now presented.  This result demonstrates that the generalized fiducial probability of the oracle graph will asymptotically dominate that of all other graphs.  Note that there is no assumption of sparsity.
\begin{theorem}[pairwise selection consistency]\label{VAR_MainResult}
Given Conditions \ref{VAR_condition1}-\ref{VAR_condition5}, for any $G \subseteq \{1, \dots, p^{2}\} \setminus G_{o}$,
\[
\frac{  r(G \mid Y)  }{  r(G_{o} \mid Y)  } \overset{P_{x}}{\longrightarrow} 0
\]
as $n \to \infty$ or $n, p \to \infty$.
\end{theorem}

If Condition \ref{VAR_condition5} is violated, Corollary \ref{VAR_MainResult_corollary} demonstrates that the generalized fiducial mass function $r(G \mid Y)$ will concentrate asymptotically on the subset of graphs $\{G : G\subseteq G_{o}\}$.  In practice, for sufficiently large $n$, this means that there will be a few graphs which the algorithm visits frequently, and the largest one (in cardinality) likely contains the greatest number of the oracle components.
\begin{corollary}[pairwise selection consistency]\label{VAR_MainResult_corollary}
Relaxing Condition \ref{VAR_condition5} in Theorem \ref{VAR_MainResult} gives, for any $G \subseteq \{1, \dots, p^{2}\} \setminus \{G : G \subseteq G_{o}\}$,
\[
\frac{  r(G \mid Y)  }{  r(G_{o} \mid Y)  } \overset{P_{x}}{\longrightarrow} 0
\]
as $n \to \infty$ or $n, p \to \infty$.
\end{corollary}

The additional corollary stated next demonstrates that the EAS methodology will concentrate all generalized fiducial mass on the true model, asymptotically, for fixed $p$.
\begin{corollary}[strong selection consistency, fixed $p$]\label{VAR_MainResult_strong_corollary}
Given Conditions \ref{VAR_condition1}-\ref{VAR_condition5} and fixed $p$,
\[
r(G_{o} \mid Y) \overset{P_{x}}{\longrightarrow} 1
\]
as $n \to \infty$.
\end{corollary}

Note the following short remark about the meaning of the difference between $pairwise$ and $strong$ model selection consistency.  The statement of $strong$ graph selection consistency is essentially a statement that the true model will be assigned large probability and all other models will be assigned small probabilities.  Conversely, the implication of $pairwise$ graph selection consistency is that the probability assigned to the true model will be large relative to each of the other model probabilities, individually, but that all models (including the true model) may have small probabilities.  Such a phenomenon is common for model selection paradigms in which the set of candidate models grows very fast with dimension (i.e., like $2^{p^{2}}$ in the case of a VAR(1) model).

The next subsection illustrates the additional attribute that our model selection consistency results are robust to model misspecification, namely, the assumption of Gaussian VAR model errors.

\subsection{Standalone supporting results}\label{VAR_standalone_results}

This subsection provides five lemmas which were foundational to our proof techniques for establishing our theory for the EAS methodology.  Non-asymptotic moment bounds on products of $\X$ and $\U$ (in the VAR model formulation (\ref{VAR_model})), with respect to $n$ and $p$, are the building blocks for any theoretical pursuit of understanding high-dimensional, multivariate VAR models.  These results are essentially a collection of second moment bounds of the quantities and cross-quantities in $
\Omega := \frac{1}{n} 
\big(\begin{smallmatrix}
\X\X' & \X\U'  \\
\U\X' & \U\U' \\
\end{smallmatrix}\big)
$, and establish the notion that our preceding fiducial consistency results will remain true under model misspecification.  This is due to the fact that as long as the VAR(1) model errors are independent within and across time and there exist bounded fourth moments (i.e., components appearing in $E_{x}(\Omega^{2})$), the following collection of lemmas will remain true (up to some constants of proportionality).  And as a consequence, our generalized fiducial consistency results (which assume Gaussian data) will still hold even if the true data is not Gaussian. 

\begin{lemma}\label{VAR_omega_bound}
Assume $\|A^{0}\|_{2} \le c$.  Then for all $n \ge \max\{N_{1},N_{2}\}$,
\[
P_{x}\Big( \big[\lambda_{\min}\big(\Omega - E_{x}(\Omega)\big)\big]^{2} > \delta^{2} \Big) 
\le V_{3},
\]
where $V_{3}$ is as in (\ref{VAR_V3}),
$
\Omega := \frac{1}{n} 
\big(\begin{smallmatrix}
\X\X' & \X\U'  \\
\U\X' & \U\U' \\
\end{smallmatrix}\big)
$,
and 
$
E_{x}(\Omega) = 
\Big(\begin{smallmatrix}
\Gamma_{n}(0) &  \\
 & I_{p} \\
\end{smallmatrix}\Big)
$.
\end{lemma}

\begin{lemma}\label{VAR_lemma_data_bound_1}
Assume $\|A^{0}\|_{2} \le c$.  Then for all $n \ge N_{1}$,
\[
\frac{1}{n^{2}}\tr\Big(E_{x}(\X\U'\U\X')\Big) \le \frac{ p (\sigma_{\max}^{0})^{2} \min\{|G_{o}|, p\} }{ n (1 - c^{2}) }.
\]
\end{lemma}

\begin{lemma}\label{VAR_lemma_data_bound_2}
Assume $\|A^{0}\|_{2} \le c$.  Then for all $n \ge N_{2}$,
\[
\tr\Big(\frac{1}{n^{2}}E_{x}\big((\X\X')^{2}\big) - \Gamma_{n}^{2}(0)\Big) \le \frac{\delta^{2}}{4}V_{2}.
\]
\end{lemma}

\begin{lemma}\label{VAR_lemma_data_bound_3}
Assume $\|A^{0}\|_{2} \le c$.  Then,
\[
\frac{1}{n^{2}}\tr\Big(E_{x}(\X\X'\X\U'A^{0})\Big) \le \frac{ 2 (\sigma_{\max}^{0})^{3} c^{2} \min\{|G_{o}|,p\}^{2}}{ (1 - c^{2})^{2} n }.
\]
\end{lemma}

\begin{lemma}\label{VAR_emp_cov_lowB}
Assume that Condition \ref{VAR_condition1} holds.  Then for all $n \ge N_{2}$,
\[
P_{x}\Big(\lambda_{\min}(\X\X'/n) \ge \delta/2\Big) \ge 1 - V_{2},
\]
where $V_{2}$ is as in (\ref{VAR_V2}).
\end{lemma}

\section{Simulation results}\label{VAR_simulation}

While the theoretical pursuits of this paper have been focused on the conditions and supporting lemmas/theorems needed for the EAS procedure to assign the highest probability to the oracle graph with probability converging to 1 as $n, p \to \infty$, we ultimately designed the EAS approach with more practical intuitions in mind.  In applications, the true data generating model, $G_{o}$, may itself contain redundant information (i.e., unnecessary active components), and through our $h$-function methodology we are able to focus on recovering only the necessary active components.  In doing so, at least for finite samples the EAS approach re-defines what is meant by the true graph.  The purpose of our asymptotic considerations was to illustrate the conditions needed for our re-defined notion of the true graph to correspond precisely to the oracle graph.  

In this section, we demonstrate on synthetic data that when the theoretical conditions are satisfied the EAS procedure performs as our asymptotic theory suggests, and is also able to perform as well as or better than existing methods in high-dimensional settings with respect to out-of-sample prediction error and estimation error.  In fact, we find and present evidence to suggest that Conditions \ref{VAR_condition1} and \ref{VAR_condition2} are useful for high-dimensional settings.  Moreover, Condition \ref{VAR_condition1} is a simple and verifiable condition for actual observed data which informs of the sample size needed for competitive performance and is so well calibrated that we demonstrate deteriorating performance when it is not satisfied.  

Furthermore, the EAS algorithm does not require any tuning parameter to achieve at or better than the out-of-sample predictive performance of competing methods such as LASSO or elastic net.  The latter, more conventional methods, require cross-validation over a grid of tuning parameters, and the appropriateness of the grid depends on the scaling of the data (i.e., $\Sigma$).  On the contrary, via our $\Lambda_{g}$ component in $\varepsilon = \Lambda_{g} \cdot \max\Big\{ 1, n^{.51}p^{2}\Big(.5\log(\log(n))|G| - |G_{o}|\Big)\Big\}$ (see (\ref{VAR_default_eps}) with $\rho = .49$) the EAS algorithm is scale invariant.  

For all of our numerical results, the component $d$ in the $h$-function is set at $d = \min_{1\le j\le p}\{m_{j}^{g_{\text{enet}}}\}/10$, where $G_{\text{enet}}$ are the active components estimated by elastic net.  And as discussed previously, the constraint $\|A_{g}\|_{2} \le c$ in the $h$-function is replaced with $\|A_{g}\|_{2} < 1$ since $c$ is not available on real data.  

In the following two subsections we present both low ($p=4, n = 120$) and high ($p=10, n=20$ and $p=30, n=180$) dimensional simulation studies on synthetic data generated according to model (\ref{VAR_DGE}).  For each of 100 random data generating seeds, the transition matrix is randomly generated according to each of the five patterns described in \cite{Han2015}.  In each instance of a transition matrix $A^{0}$ the $p$ diagonal components are active, and for patterns with additional randomly assigned active/inactive components the probability of each component being generated as active is .01.  Values of each diagonal component are assigned by sampling from the $N(\pm12,1)$ distribution, while off-diagonal component values are assigned by sampling from the $N(\pm3,1)$ distribution.  As is common practice (e.g., \citealt{Han2015}), after a given $A^{0}$ is randomly generated it is rescaled so that $\|A^{0}\|_{2} = .5 =: c$, and as in \cite{Han2015} the contemporaneous error covariance matrix $\Sigma^{0} := I_{p}$.

In all simulation designs, the performance of the EAS algorithm is compared to that of LASSO and elastic net implementations, and to a recent ``direct estimation of high-dimensional stationary VAR'' estimation procedure proposed by \cite{Han2015} which is formulated as a linear program (we denote this procedure by DELP for ``direct estimation linear program'').  The LASSO and elastic net routines are implemented from the $Python$ module \verb1scikit-learn1 \cite{scikit-learn}, along with their builtin cross-validation procedures for time-series data.  For the DELP routine, the authors of \cite{Han2015} were kind enough to provide their $R$ code.  However, we had to supplement their provided code by writing code to implement the cross-validation procedure they propose in \cite{Han2015} for selecting their tuning parameter.  Note that we generate synthetic data consistent with that described in \cite{Han2015} so that the scaling of the data is appropriate for their default grid of tuning parameters for cross-validation.

The entirety of the simulation study was computed in parallel on a computing cluster, and completed in approximately one day of run time.  The code/workflow for reproducing all numerical results presented in this paper can be found at \verb1https://jonathanpw.github.io/research1.

\subsection{Definitions of performance metrics}\label{VAR_def_metrics}

A variety of metrics are considered for evaluating performance across procedures.  For each random generator seed for each simulation design, $2 n$ instances of the time-series are generated with $X^{(0)} = 0_{p\times1}$.  The first $n$ are used for estimation, and the last $n$ are set aside as an out-of-sample test set.  As in \cite{Han2015}, on the out-of-sample test set we compute the $L_{2}$ prediction error, $\frac{1}{n}\|\Y - \widehat{A}\X\|_{2}$, and the $L_{F}$ prediction error, $\frac{1}{n}\|\Y - \widehat{A}\X\|_{F}$, where $\widehat{A}$ represents the estimated transition matrix on the first $n$, in-sample, time instances.  As in \cite{Basu2015} and \cite{Ghosh2018}, we also calculate the estimation error, $\|\widehat{A} - A^{0}\|_{F}/\|A^{0}\|_{F}$.  For the EAS procedure, $\widehat{A}$ is computed analogously to Bayesian model averaging, with least squares estimates used for every visited graph in the MCMC chain.

Additionally, we report $|G_{\text{MAP}}|$ as the number of nonzero (or active) components in the estimated graph for the frequentist LASSO, elastic net, and DELP procedures, and as the number of active components in the most frequently visited graph (i.e., maximum a-posteriori probability or MAP) for the MCMC-based EAS algorithm.  The false positive rate (FPR) is computed as the number of the $p^{2}$ components in the estimated transition matrix incorrectly set active, as a proportion of the number of truly inactive components.  Conversely, the false negative rate (FNR) is computed as the number of the $p^{2}$ components in the estimated transition matrix incorrectly set inactive, as a proportion of the number of truly active components.  For the EAS procedure, the FPR and FNR are computed based on the estimated $G_{\text{MAP}}$.  

\subsection{Low-dimensional setting}\label{VAR_lowD_sim}

This first simulation design serves to demonstrate that the EAS procedure performs consistently with what the theory in Section \ref{VAR_theory} suggests for data with $p^{2} < n$.  For this simulation we present two additional performance metrics, $\widehat{r}(G_{o} \mid Y)$ and $\#\{G_{\text{MAP}}=G_{o}\}$.  The former is the estimated generalized fiducial probability of the oracle model, calculated as the number of times the MCMC algorithm visited $G_{o}$ divided by the number of steps of the chain.  This metric is only available within the EAS framework because relative model probabilities are computed.  The latter metric, $\#\{G_{\text{MAP}}=G_{o}\}$, is the proportion, over all 100 generated data sets, of instances in which the estimated $G_{\text{MAP}}$ corresponds precisely to $G_{o}$.

\begin{table}[H]
\footnotesize
\centering
\begin{tabular}{c | ccccc}
\multicolumn{6}{c}{Random pattern transition matrix} \\
& \multicolumn{5}{c}{p = 4, n = 120} \\ 
& oracle & eas & delp & lasso & enet \\
\hline
L2&1.27&1.28&1.31&1.3&1.3\\
&(0.11)&(0.11)&(0.12)&(0.12)&(0.12)\\
\hline
LF&2.01&2.02&2.03&2.03&2.03\\
&(0.07)&(0.07)&(0.07)&(0.07)&(0.07)\\
\hline
est err&0.17&0.21&0.32&0.32&0.33\\
&(0.06)&(0.1)&(0.11)&(0.1)&(0.1)\\
\hline
$|G_{\text{MAP}}|$&4.12&4.0&7.84&7.94&8.39\\
&(0.35)&(0.32)&(3.34)&(2.97)&(3.24)\\
\hline
FPR&&0.01&0.32&0.33&0.36\\
&&(0.02)&(0.28)&(0.25)&(0.27)\\
\hline
FNR&&0.04&0.01&0.01&0.01\\
&&(0.09)&(0.05)&(0.05)&(0.05)\\
\hline
$\widehat{r}(G_{o}\mid Y)$&&0.7&&&\\
&&(0.32)&&&\\
\hline
$\#\{G_{\text{MAP}}=G_{o}\}$&&0.81&0.08&0.11&0.11\\
\hline
& \multicolumn{5}{c}{r.h.s. Condition \ref{VAR_condition1} = 10.10 (s.e. 0.90) vs 5} \\
& \multicolumn{5}{c}{prop data sets Condition \ref{VAR_condition2} satisfied = 0.83} \\
\end{tabular}\caption{\footnotesize See Section \ref{VAR_def_metrics} for definitions of each performance metric, except for the last two which are described in Section \ref{VAR_lowD_sim}.  All metrics are quantities averaged over 100 generated data sets, and standard errors are in parentheses.  The `oracle' column displays corresponding characteristics in the case that the oracle graph, $G_{o}$, is known, and using the least squares estimate of $A^{0}$.  Note that for Condition \ref{VAR_condition1}, $4(1+c^{2}) = 5$.  Recall that a new set of active components $G_{o}$ are generated for each data set, which gives the variability for $|G_{\text{MAP}}|$ in the `oracle' column.}\label{VAR_lowD_table}  
\end{table}

Observe from Table \ref{VAR_lowD_table} that the EAS procedure performs very competitively with these existing methods; better average performance metric values across the board, but all routines are within about one standard error of each other.  Furthermore, the EAS algorithm selected a $G_{\text{MAP}}$ with 3-4 fewer active components, on average, with $G_{\text{MAP}} = G_{o}$ for 81 of the 100 of the data sets.  This is far better graph selection than the competing methods which consistently over-select active components.  Note that based on the proportion of data sets in which Condition \ref{VAR_condition2} is satisfied, the oracle model is only identifiable for the EAS algorithm in 83 percent of the data sets.  In other words, our theory would suggest that the EAS procedure should identify the true model in 83 of the 100 data sets considered, and in actuality the EAS algorithm identified the true model in 81 of the 100 data sets.

\subsection{High-dimensional setting}

The tables in this section display the results of two high-dimensional simulation designs in which $p^{2} > n$, and for all five transition matrix patterns.  

\begin{table}[H]
\footnotesize
\centering
\begin{tabular}{c | ccccc | ccccc}
\multicolumn{11}{c}{Band pattern transition matrix} \\
& \multicolumn{5}{c|}{p = 10, n = 20} & \multicolumn{5}{c}{p = 30, n = 180} \\ 
& oracle & eas & delp & lasso & enet & oracle & eas & delp & lasso & enet \\
\hline
L2&3.04&3.29&4.13&2.94&2.9&1.92&2.02&2.03&2.02&2.02\\
&(0.68)&(0.65)&(5.9)&(0.49)&(0.45)&(0.09)&(0.09)&(0.1)&(0.1)&(0.1)\\
\hline
LF&3.46&3.54&3.55&3.4&3.39&5.53&5.64&5.67&5.65&5.65\\
&(0.22)&(0.24)&(0.71)&(0.18)&(0.18)&(0.06)&(0.06)&(0.07)&(0.07)&(0.07)\\
\hline
est err&1.07&1.24&1.15&0.97&0.94&0.34&0.63&0.68&0.64&0.65\\
&(0.17)&(0.18)&(0.77)&(0.04)&(0.05)&(0.03)&(0.06)&(0.05)&(0.06)&(0.06)\\
\hline
$|G_{\text{MAP}}|$&28.0&11.64&8.89&3.43&19.63&88.0&22.32&43.43&49.78&60.85\\
&(0.0)&(2.88)&(23.38)&(4.56)&(18.44)&(0.0)&(2.92)&(25.39)&(11.55)&(29.62)\\
\hline
FPR&&0.1&0.08&0.02&0.17&&0.0&0.01&0.02&0.03\\
&&(0.03)&(0.24)&(0.04)&(0.18)&&(0.0)&(0.03)&(0.01)&(0.03)\\
\hline
FNR&&0.83&0.88&0.93&0.74&&0.75&0.63&0.59&0.57\\
&&(0.06)&(0.23)&(0.07)&(0.21)&&(0.03)&(0.05)&(0.05)&(0.07)\\
\hline
& \multicolumn{5}{c|}{r.h.s. Condition \ref{VAR_condition1} = 0.7112 (s.e. 0.2384) vs 5} & \multicolumn{5}{c}{r.h.s. Condition \ref{VAR_condition1} = 5.7756 (s.e. 0.4147) vs 5}  \\
& \multicolumn{5}{c|}{prop data sets Condition \ref{VAR_condition2} satisfied = 0} & \multicolumn{5}{c}{prop data sets Condition \ref{VAR_condition2} satisfied = 0} \\
\end{tabular}\caption{\footnotesize See caption for Table \ref{VAR_lowD_table}.}\label{VAR_highD_table_band}  
\end{table}

An important distinction to observe between the two designs, for all transition matrix patterns, is that for the $p = 10, n = 20$ case Condition \ref{VAR_condition1} is never satisfied, while it is always satisfied for the $p = 30, n = 180$ case.  This occurrence is by design to demonstrate the deteriorated performance of the EAS algorithm when this important, well-calibrated, and verifiable condition is not satisfied.  In the $p = 30, n = 180$ case the EAS algorithm performs just as well, or better than the competing methods, with respect to all metrics.

\begin{table}[H]
\footnotesize
\centering
\begin{tabular}{c | ccccc | ccccc}
\multicolumn{11}{c}{Cluster pattern transition matrix} \\
& \multicolumn{5}{c|}{p = 10, n = 20} & \multicolumn{5}{c}{p = 30, n = 180} \\ 
& oracle & eas & delp & lasso & enet & oracle & eas & delp & lasso & enet \\
\hline
L2&2.66&3.52&10.35&3.22&3.16&1.91&1.96&2.03&2.01&2.01\\
&(0.37)&(0.85)&(44.59)&(0.51)&(0.48)&(0.1)&(0.12)&(0.11)&(0.11)&(0.11)\\
\hline
LF&3.28&3.6&4.05&3.51&3.48&5.5&5.55&5.63&5.61&5.61\\
&(0.16)&(0.24)&(2.06)&(0.18)&(0.18)&(0.06)&(0.07)&(0.07)&(0.06)&(0.06)\\
\hline
est err&0.48&1.08&1.42&0.95&0.92&0.17&0.34&0.5&0.46&0.46\\
&(0.12)&(0.16)&(1.49)&(0.05)&(0.06)&(0.03)&(0.09)&(0.05)&(0.05)&(0.05)\\
\hline
$|G_{\text{MAP}}|$&10.39&12.16&18.51&4.33&21.05&31.24&27.64&42.28&47.76&48.14\\
&(0.68)&(2.28)&(34.53)&(5.52)&(19.45)&(1.26)&(2.47)&(23.75)&(8.28)&(8.4)\\
\hline
FPR&&0.09&0.17&0.03&0.18&&0.0&0.01&0.02&0.02\\
&&(0.03)&(0.35)&(0.04)&(0.19)&&(0.0)&(0.03)&(0.01)&(0.01)\\
\hline
FNR&&0.63&0.68&0.8&0.55&&0.14&0.04&0.04&0.04\\
&&(0.14)&(0.33)&(0.19)&(0.29)&&(0.08)&(0.04)&(0.04)&(0.04)\\
\hline
& \multicolumn{5}{c|}{r.h.s. Condition \ref{VAR_condition1} = 0.6941 (s.e. 0.2265) vs 5} & \multicolumn{5}{c}{r.h.s. Condition \ref{VAR_condition1} = 6.0315 (s.e. 0.4613) vs 5}  \\
& \multicolumn{5}{c|}{prop data sets Condition \ref{VAR_condition2} satisfied = 0} & \multicolumn{5}{c}{prop data sets Condition \ref{VAR_condition2} satisfied = 0} \\
\end{tabular}\caption{\footnotesize See caption for Table \ref{VAR_lowD_table}.  Recall that a new set of active components $G_{o}$ are generated for each data set, which gives the variability for $|G_{\text{MAP}}|$ in the `oracle' column.}\label{VAR_highD_table_cluster}  
\end{table}

\begin{table}[H]
\footnotesize
\centering
\begin{tabular}{c | ccccc | ccccc}
\multicolumn{11}{c}{Hub pattern transition matrix} \\
& \multicolumn{5}{c|}{p = 10, n = 20} & \multicolumn{5}{c}{p = 30, n = 180} \\ 
& oracle & eas & delp & lasso & enet & oracle & eas & delp & lasso & enet \\
\hline
L2&3.06&3.3&6.52&3.02&2.98&1.93&2.03&2.05&2.03&2.03\\
&(0.64)&(0.66)&(29.28)&(0.61)&(0.58)&(0.09)&(0.09)&(0.1)&(0.1)&(0.1)\\
\hline
LF&3.44&3.54&3.68&3.41&3.4&5.53&5.63&5.66&5.64&5.65\\
&(0.21)&(0.2)&(1.53)&(0.2)&(0.2)&(0.06)&(0.06)&(0.06)&(0.06)&(0.06)\\
\hline
est err&1.06&1.23&1.26&0.98&0.96&0.33&0.63&0.69&0.65&0.66\\
&(0.17)&(0.17)&(1.15)&(0.03)&(0.05)&(0.03)&(0.05)&(0.05)&(0.06)&(0.05)\\
\hline
$|G_{\text{MAP}}|$&26.0&11.95&11.9&2.89&19.74&78.0&21.97&41.24&47.64&61.35\\
&(0.0)&(2.48)&(29.42)&(4.84)&(20.78)&(0.0)&(2.72)&(24.41)&(11.38)&(32.45)\\
\hline
FPR&&0.1&0.11&0.02&0.17&&0.0&0.01&0.02&0.03\\
&&(0.03)&(0.3)&(0.04)&(0.2)&&(0.0)&(0.03)&(0.01)&(0.03)\\
\hline
FNR&&0.82&0.86&0.94&0.73&&0.73&0.6&0.57&0.54\\
&&(0.07)&(0.29)&(0.09)&(0.25)&&(0.03)&(0.05)&(0.05)&(0.07)\\
\hline
& \multicolumn{5}{c|}{r.h.s. Condition \ref{VAR_condition1} = 0.7 (s.e. 0.2323) vs 5} & \multicolumn{5}{c}{r.h.s. Condition \ref{VAR_condition1} = 5.667 (s.e. 0.4128) vs 5}  \\
& \multicolumn{5}{c|}{prop data sets Condition \ref{VAR_condition2} satisfied = 0} & \multicolumn{5}{c}{prop data sets Condition \ref{VAR_condition2} satisfied = 0} \\
\end{tabular}\caption{\footnotesize See caption for Table \ref{VAR_lowD_table}.}\label{VAR_highD_table_hub}  
\end{table}

Notice also that the high-dimensional numerical results presented in this section do not list $\widehat{r}(G_{o} \mid Y)$ nor $\#\{G_{\text{MAP}}=G_{o}\}$ as performance metrics.  For each of the estimation methods, the metric $\#\{G_{\text{MAP}}=G_{o}\}$ (and $\widehat{r}(G_{o} \mid Y)$ for EAS) produces zeros in almost all cases.  For the EAS algorithm, this is due to the fact that Condition \ref{VAR_condition2} is never satisfied for these high-dimensional simulation designs, and so the oracle model is $not$ identifiable for the EAS procedure.  However, we would not necessarily expect Condition \ref{VAR_condition2} to be satisfied when $p^{2} > n$ since our theory does not apply.

\begin{table}[H]
\footnotesize
\centering
\begin{tabular}{c | ccccc | ccccc}
\multicolumn{11}{c}{Random pattern transition matrix} \\
& \multicolumn{5}{c|}{p = 10, n = 20} & \multicolumn{5}{c}{p = 30, n = 180} \\ 
& oracle & eas & delp & lasso & enet & oracle & eas & delp & lasso & enet \\
\hline
L2&2.6&3.36&5.21&3.06&3.03&1.91&1.99&2.03&2.01&2.01\\
&(0.4)&(0.8)&(9.99)&(0.49)&(0.5)&(0.09)&(0.1)&(0.1)&(0.1)&(0.1)\\
\hline
LF&3.24&3.55&3.71&3.45&3.43&5.51&5.58&5.64&5.62&5.62\\
&(0.17)&(0.26)&(0.95)&(0.21)&(0.22)&(0.05)&(0.06)&(0.06)&(0.06)&(0.06)\\
\hline
est err&0.49&1.07&1.2&0.95&0.92&0.2&0.44&0.55&0.51&0.51\\
&(0.15)&(0.16)&(0.82)&(0.06)&(0.07)&(0.03)&(0.08)&(0.05)&(0.05)&(0.05)\\
\hline
$|G_{\text{MAP}}|$&10.9&12.49&14.51&3.69&19.28&38.72&25.65&45.45&50.07&50.55\\
&(0.96)&(2.68)&(30.43)&(4.68)&(20.07)&(3.13)&(2.68)&(30.98)&(9.95)&(10.89)\\
\hline
FPR&&0.1&0.13&0.02&0.17&&0.0&0.02&0.02&0.02\\
&&(0.03)&(0.31)&(0.03)&(0.19)&&(0.0)&(0.03)&(0.01)&(0.01)\\
\hline
FNR&&0.63&0.72&0.82&0.58&&0.35&0.21&0.19&0.19\\
&&(0.15)&(0.31)&(0.18)&(0.31)&&(0.09)&(0.07)&(0.07)&(0.07)\\
\hline
& \multicolumn{5}{c|}{r.h.s. Condition \ref{VAR_condition1} = 0.6961 (s.e. 0.2523) vs 5} & \multicolumn{5}{c}{r.h.s. Condition \ref{VAR_condition1} = 5.7805 (s.e. 0.43) vs 5}  \\
& \multicolumn{5}{c|}{prop data sets Condition \ref{VAR_condition2} satisfied = 0} & \multicolumn{5}{c}{prop data sets Condition \ref{VAR_condition2} satisfied = 0} \\
\end{tabular}\caption{\footnotesize See caption for Table \ref{VAR_lowD_table}.  Recall that a new set of active components $G_{o}$ are generated for each data set, which gives the variability for $|G_{\text{MAP}}|$ in the `oracle' column.}\label{VAR_highD_table_random}  
\end{table}

\begin{table}[H]
\footnotesize
\centering
\begin{tabular}{c | ccccc | ccccc}
\multicolumn{11}{c}{Scale--free pattern transition matrix} \\
& \multicolumn{5}{c|}{p = 10, n = 20} & \multicolumn{5}{c}{p = 30, n = 180} \\ 
& oracle & eas & delp & lasso & enet & oracle & eas & delp & lasso & enet \\
\hline
L2&3.37&3.19&8.31&2.86&2.83&1.94&2.0&2.01&2.01&2.0\\
&(0.81)&(0.75)&(30.93)&(0.47)&(0.47)&(0.09)&(0.1)&(0.1)&(0.1)&(0.1)\\
\hline
LF&3.51&3.49&3.83&3.36&3.35&5.53&5.61&5.62&5.63&5.62\\
&(0.22)&(0.22)&(1.73)&(0.19)&(0.18)&(0.06)&(0.06)&(0.06)&(0.07)&(0.06)\\
\hline
est err&1.36&1.32&1.56&0.99&0.96&0.52&0.84&0.87&0.9&0.86\\
&(0.27)&(0.19)&(1.76)&(0.03)&(0.04)&(0.04)&(0.04)&(0.04)&(0.06)&(0.05)\\
\hline
$|G_{\text{MAP}}|$&28.0&11.8&15.88&2.07&16.97&88.0&15.14&23.22&17.66&77.21\\
&(0.0)&(2.85)&(34.04)&(3.76)&(18.88)&(0.0)&(2.24)&(4.07)&(10.06)&(35.83)\\
\hline
FPR&&0.1&0.15&0.01&0.15&&0.0&0.01&0.0&0.06\\
&&(0.03)&(0.34)&(0.03)&(0.18)&&(0.0)&(0.0)&(0.0)&(0.03)\\
\hline
FNR&&0.84&0.82&0.96&0.78&&0.86&0.79&0.83&0.63\\
&&(0.07)&(0.34)&(0.07)&(0.22)&&(0.03)&(0.04)&(0.09)&(0.11)\\
\hline
& \multicolumn{5}{c|}{r.h.s. Condition \ref{VAR_condition1} = 0.6866 (s.e. 0.2609) vs 5} & \multicolumn{5}{c}{r.h.s. Condition \ref{VAR_condition1} = 5.3767 (s.e. 0.4072) vs 5}  \\
& \multicolumn{5}{c|}{prop data sets Condition \ref{VAR_condition2} satisfied = 0} & \multicolumn{5}{c}{prop data sets Condition \ref{VAR_condition2} satisfied = 0} \\
\end{tabular}\caption{\footnotesize See caption for Table \ref{VAR_lowD_table}.}\label{VAR_highD_table_scalefree}  
\end{table}

Moreover, recall that in finite samples, and particularly high-dimensional, settings with highly-correlated data the EAS framework was developed with the intuition that the oracle graph itself may not be $\varepsilon$-$admissible$.  In these settings, the EAS methodology re-defines the notion of the `true' graph to be some non-redundant subgraph of the oracle graph, at least non-asymptotically.  This is validated empirically in the tables that follow by observing that when Condition \ref{VAR_condition1} is satisfied the EAS algorithm almost always requires fewer active components to achieve on par or better performance than the competing methods, with respect to all metrics.  Recall also that the EAS algorithm has no tuning parameter, while the competing methods use cross-validation to optimize out-of-sample prediction accuracy.

\section{Real data application}\label{VAR_real_data}

As a final exposition of the EAS methodology developed for the VAR(1) model, this section presents results of implementing the algorithm on real data.  

\begin{figure}[H]
\centering
\includegraphics[scale=.5]{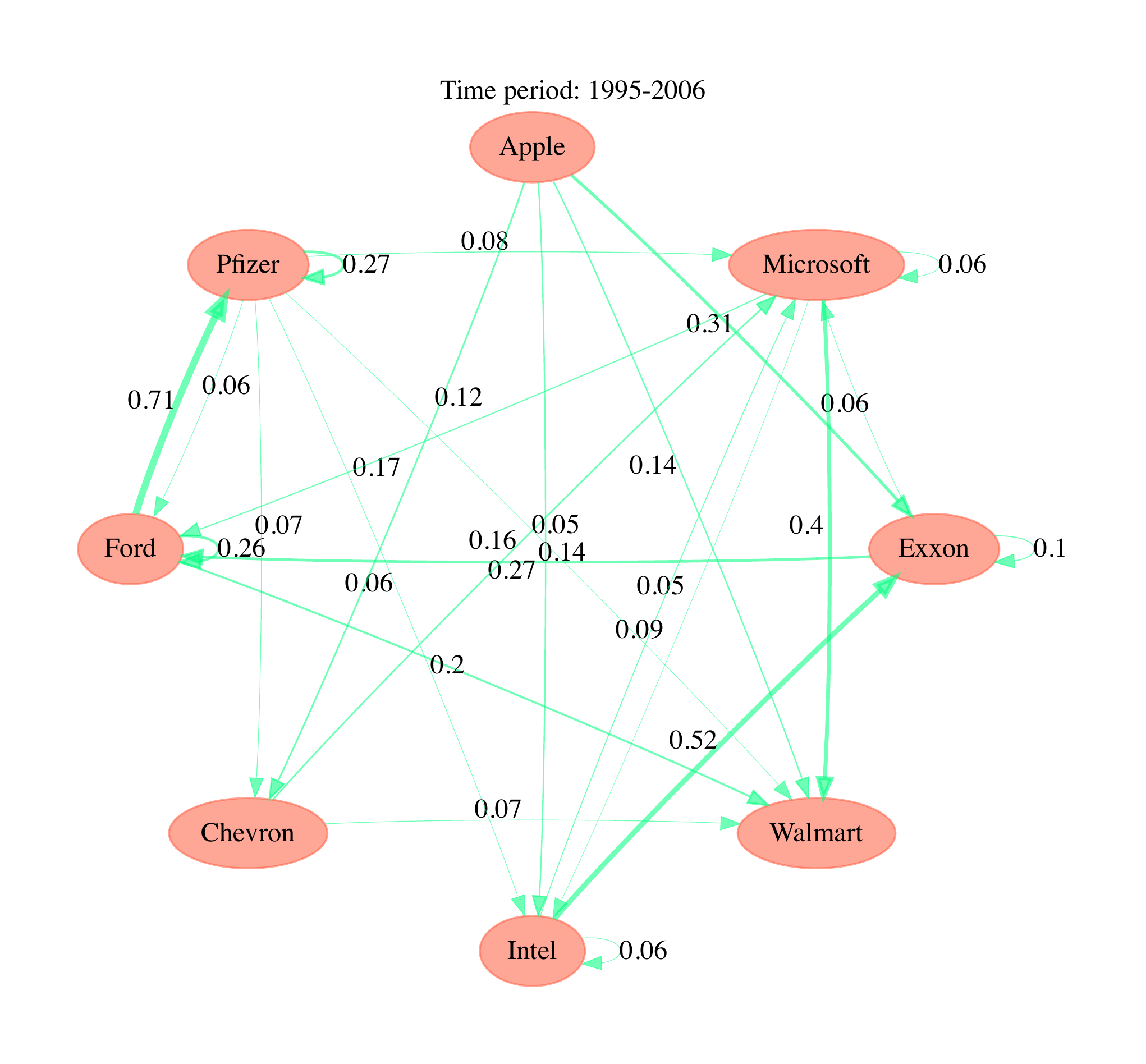}
\caption{\footnotesize Directed graph of inclusion probabilities of components of the transition matrix, $A$, for monthly closing stock price of 8 companies.  First differences of the data are used.  Each edge label represents the marginal generalized fiducial (or posterior-like) inclusion probability of a particular component of $A$.  That is, the proportion of graphs, $G$, (over all MCMC-sampled graphs) in which each component (i.e., edge) of $A$ is active.  Line widths are proportional to inclusion probabilities, and inclusion probabilities less than .05 are omitted.}\label{VAR_stocks_1995_2006}
\end{figure}

\begin{figure}[H]
\centering
\includegraphics[scale=.5]{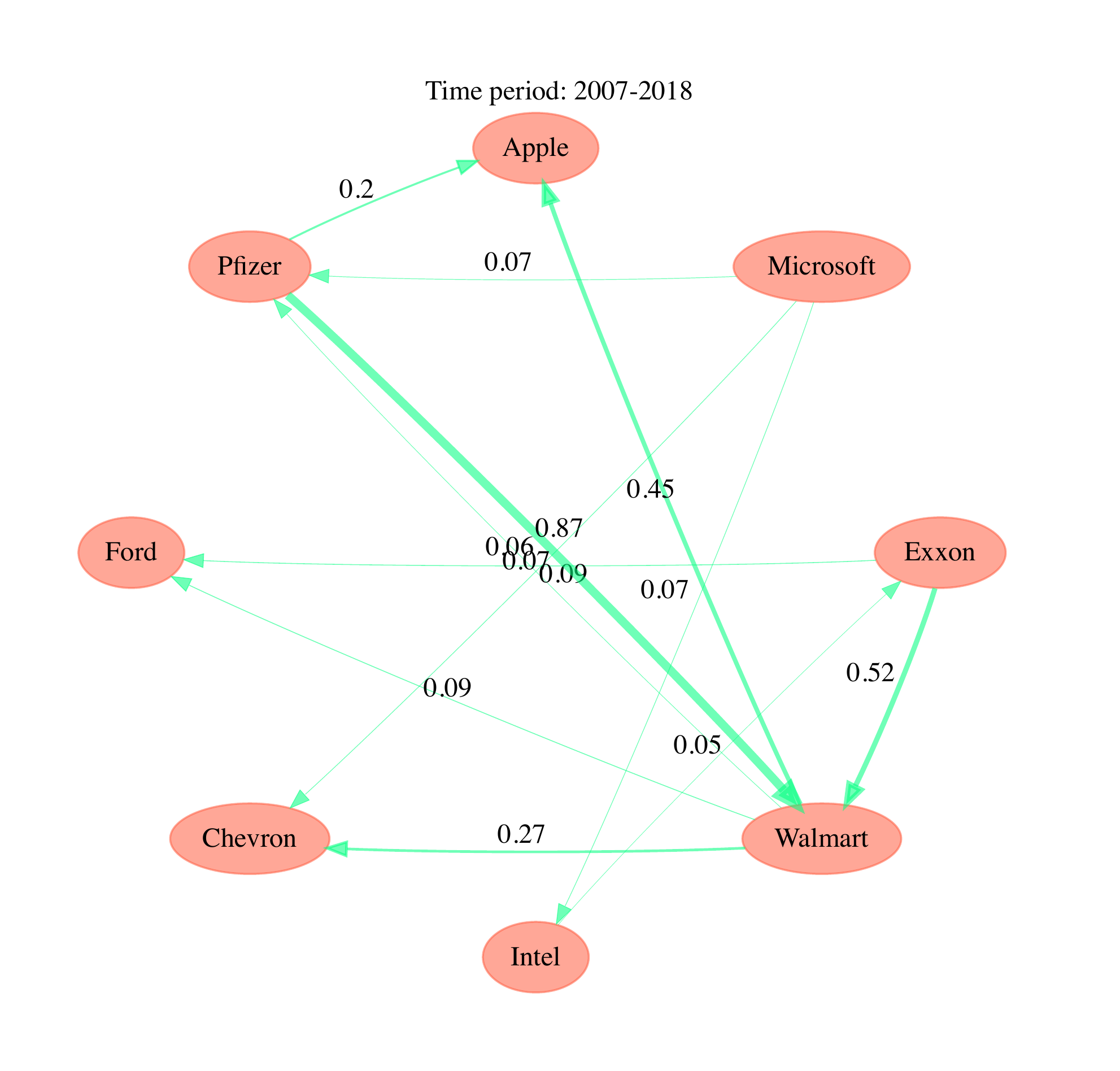}
\caption{\footnotesize See description for Figure \ref{VAR_stocks_1995_2006}.}\label{VAR_stocks_2007_2018}
\end{figure}

Monthly closing stock price data for eight well-known companies from 1995-2018 are downloaded from Yahoo Finance via the R package \verb1BatchGetSymbols1 \citep{BatchGetSymbols}.  First-differences of the time-series are used for stationarity, and the data is split into two time periods, 1995-2006 and 2007-2018.  It is verified that Condition \ref{VAR_condition1} is satisfied for the time period 2007-2018 (9.33 versus $8 = 4(1 + 1^{2})$), but not for 1995-2006 (2.05 versus $8 = 4(1 + 1^{2})$).  This occurrence is useful for observing the performance of the EAS procedure on real data when the condition is and is not satisfied.  

The results are displayed graphically in Figures \ref{VAR_stocks_1995_2006} and \ref{VAR_stocks_2007_2018}.  Nodes represent individual company stocks, and each edge label represents the marginal generalized fiducial (or posterior-like) inclusion probability of a particular component of $A$.  That is the proportion of graphs, $G$, (over all MCMC-sampled graphs) in which each component (i.e., edge) of $A$ is active.  Line widths are proportional to inclusion probabilities, and inclusion probabilities less than .05 are omitted.  

Interpretation of the findings on these data should be restricted to the time period 2007-2018 in which Condition \ref{VAR_condition1} is satisfied.  However, the real data analysis conducted here is not a thorough investigation of these time-series, but rather a ``proof of concept'' for how the EAS methodology can be useful on real data.  A well qualified study would require considerable additional analysis of the data which is beyond the scope of our paper.

Nonetheless, the results do appear sensible.  From Table \ref{VAR_stocks_2007_2018}, it is observed that seemingly redundant time-series in the system such as for the two oil companies, Chevron and Exxon, do not have simultaneous marginal inclusions to a large extent, and the system is dominated by relatively few strong links.  Since many of the considered stocks correspond to consumer goods corporations, it is reasonable that the results suggest the system has numerous links to and from the massive retailer Walmart, with an especially high link from the pharmaceutical giant Pfizer.  Additionally, we see that the somewhat surprisingly strong link in Figure \ref{VAR_stocks_1995_2006} between what we would suspect are unrelated corporations/stock prices, Ford and Pfizer, vanishes in Figure \ref{VAR_stocks_2007_2018}.

Note that such a graphical representation of the results, with marginal inclusion probabilities for all components of $A$, is not possible via frequentist nor Bayesian point estimation based procedures.  This is a major advantage of estimating relative model probabilities (i.e., $r(G\mid Y)$) versus simply coefficients.  MCMC-based approaches are computationally more expensive, but they provide more information for uncertainty quantification.  The code/workflow for obtaining the real data and reproducing these result can be found at \verb1https://jonathanpw.github.io/research1.

\section{Concluding remarks}

In summary, while BVAR models have been developed and explored empirically (primarily in the econometrics literature) there exist very few theoretical investigations of the repeated sampling properties for BVAR models in the literature.  To the best of our knowledge, our established $pairwise$ and $strong$ model selection consistency results are the first of their kind in the BVAR literature.  These types of results are sure to be followed by similar results in the high-dimensional BVAR literature, analogous to the emergence of model selection strong consistency results in the high-dimensional Bayesian linear regression literature such as \cite{Johnson2012, Narisetty2014, Williams2019}.

All things considered, while it is required for our theory that $n$ exceeds some polynomial of $p$, consistent with our survey of the literature, it is claimed in \cite{Ghosh2018} that general posterior consistency results are not available for ``large p small n'' settings.  Furthermore, our graphical selection consistency results provide a theoretical guarantee for model selection, which is stronger than establishing estimation consistency of a point estimator of the VAR model parameters, and our theory is robust to model misspecification.

Moreover, recall that in finite samples, and particularly high-dimensional, settings with highly-correlated data the EAS framework was developed with the intuition that the oracle graph itself may not be $\varepsilon$-$admissible$.  In these settings, the EAS methodology re-defines the notion of the `true' graph to be some non-redundant subgraph of the oracle graph, at least non-asymptotically.  Accordingly, with our EAS methodology, we hope to demonstrate the idea that to develop inherently scalable methodology the key may be to re-think what one should hope to recover for useful statistical inference from a data generating model.


\begin{supplement}[id=suppA]
  \sname{Supplement to}
  \stitle{``The EAS approach for graphical selection consistency in vector autoregression models''}
  \slink[doi]{10.1214/00-AOASXXXXSUPP}
  \sdatatype{.pdf}
  \sdescription{For conciseness of the manuscript, longer derivations, additional lemmas, and proofs have been moved to these supplementary material.}
\end{supplement}

\bibliographystyle{imsart-nameyear}
\bibliography{References}

\section{Appendix}
This section provides proofs of the main theorem and its corollaries.  See the supplementary material for proofs of all other results.


{\noindent \bf Proof of Theorem \ref{VAR_MainResult}.}
Assume throughout this proof that $n \ge \max\{N_{1},N_{2},N_{3}\}$ (see Definition \ref{VAR_def_quantities}).  From (\ref{VAR_pmf}),
\[
\begin{split}
\frac{  r(G \mid Y)  }{  r(G_{o} \mid Y)  } 
& = (2\pi)^{\frac{|G| - |G_{o}|}{2}} n^{\frac{|G_{o}|-|G|}{2}} \frac{   E\Big( h\big(\balpha_{G}, \{\sigma_{j}\}\big)|\widetilde{\D}'_{g}\widetilde{\D}_{g}|^{\frac{1}{2}} \Big)    }{   E\Big( h\big(\balpha_{G_{o}}, \{\sigma_{j}\}\big)|\widetilde{\D}'_{g_{o}}\widetilde{\D}_{g_{o}}|^{\frac{1}{2}} \Big)   } \\
& \hspace{.25in} \times  \prod_{j=1}^{p}\Bigg[\frac{     \big|(\X\X')_{r_{j}^{g_{o}},r_{j}^{g_{o}}}\big|^{\frac{1}{2}}   }{    \big|(\X\X')_{r_{j}^{g},r_{j}^{g}}\big|^{\frac{1}{2}}    }  \cdot \frac{\big(\frac{m_{j}^{g}}{2}\big)^{-\frac{n-|r_{j}^{g}|}{2}}   }{    \big(\frac{m_{j}^{g_{o}}}{2}\big)^{-\frac{n-|r_{j}^{g_{o}}|}{2}}     }  \cdot \frac{\Gamma\Big(\frac{n-|r_{j}^{g}|}{2}\Big)    }{       \Gamma\Big(\frac{n-|r_{j}^{g_{o}}|}{2}\Big)   }\Bigg]. \\
\end{split}
\]
From \cite{Jameson2013},
\[
\frac{\Gamma\Big(\frac{n-|r_{j}^{g}|}{2}\Big)    }{       \Gamma\Big(\frac{n-|r_{j}^{g_{o}}|}{2}\Big)   } \le
\begin{cases}
\Big(\frac{n-|r_{j}^{g_{o}}|}{2}\Big)\Big(\frac{n-|r_{j}^{g}|}{2}\Big)^{\frac{|r_{j}^{g_{o}}|-|r_{j}^{g}|}{2} - 1} & \text{ if } |r_{j}^{g_{o}}|-|r_{j}^{g}| \ge 1 \\
1 & \text{ if } |r_{j}^{g_{o}}|-|r_{j}^{g}| = 0 \\
\Big(\frac{n-|r_{j}^{g}|}{2} - 1\Big)^{\frac{-(|r_{j}^{g}|-|r_{j}^{g_{o}}|)}{2}} & \text{ if } |r_{j}^{g_{o}}|-|r_{j}^{g}| \le -1 \\
\end{cases}, \\
\]
and so for $n - p \ge 4$,
\[
\begin{split}
\prod_{j=1}^{p}\frac{\Gamma\Big(\frac{n-|r_{j}^{g}|}{2}\Big)    }{       \Gamma\Big(\frac{n-|r_{j}^{g_{o}}|}{2}\Big)   } & \le \prod_{j=1}^{p} 
\begin{cases}
\big(\frac{n}{2}\big)^{\frac{|r_{j}^{g_{o}}|-|r_{j}^{g}|}{2}} & \text{ if } |r_{j}^{g_{o}}|-|r_{j}^{g}| \ge 3 \\
\frac{n}{2} & \text{ if } |r_{j}^{g_{o}}|-|r_{j}^{g}| = 2 \\
\frac{n}{2}\big(\frac{n-p}{2}\big)^{-\frac{1}{2}} & \text{ if } |r_{j}^{g_{o}}|-|r_{j}^{g}| = 1 \\
1 & \text{ if } |r_{j}^{g_{o}}|-|r_{j}^{g}| = 0 \\
\big(\frac{n-p}{2} - 1\big)^{\frac{-(|r_{j}^{g}|-|r_{j}^{g_{o}}|)}{2}} & \text{ if } |r_{j}^{g_{o}}|-|r_{j}^{g}| \le -1 \\
\end{cases} \\
& \le \prod_{j=1}^{p}\Big(\frac{n}{2}\Big)^{\max\big\{\frac{|r_{j}^{g_{o}}|}{2},1\big\}} \\
& \le \Big(\frac{n}{2}\Big)^{\frac{|G_{o}|}{2} + p}. \\
\end{split}
\]
This bound, together with the simplification,
\[
\prod_{j=1}^{p}\frac{\big(\frac{m_{j}^{g}}{2}\big)^{-\frac{n-|r_{j}^{g}|}{2}}   }{    \big(\frac{m_{j}^{g_{o}}}{2}\big)^{-\frac{n-|r_{j}^{g_{o}}|}{2}}     } 
= 2^{\frac{|G_{o}| - |G|}{2}}   \prod_{j=1}^{p}  \frac{(m_{j}^{g_{o}})^{\frac{n-|r_{j}^{g_{o}}|}{2}}}{(m_{j}^{g})^{\frac{n-|r_{j}^{g}|}{2}}} 
\]
gives
\[
\begin{split}
\frac{  r(G \mid Y)  }{  r(G_{o} \mid Y)  } & \le \Big(\frac{\pi}{n}\Big)^{\frac{|G| - |G_{o}|}{2}} \Big(\frac{n}{2}\Big)^{\frac{|G_{o}|}{2} + p} \frac{   E\Big( h\big(\balpha_{G}, \{\sigma_{j}\}\big)|\widetilde{\D}'_{g}\widetilde{\D}_{g}|^{\frac{1}{2}} \Big)    }{   E\Big( h\big(\balpha_{G_{o}}, \{\sigma_{j}\}\big)|\widetilde{\D}'_{g_{o}}\widetilde{\D}_{g_{o}}|^{\frac{1}{2}} \Big)   } \\
& \hspace{.25in} \times  \prod_{j=1}^{p}\Bigg[\frac{     \big|(\X\X')_{r_{j}^{g_{o}},r_{j}^{g_{o}}}\big|^{\frac{1}{2}}   }{    \big|(\X\X')_{r_{j}^{g},r_{j}^{g}}\big|^{\frac{1}{2}}    }  \cdot \frac{(m_{j}^{g_{o}})^{\frac{n-|r_{j}^{g_{o}}|}{2}}}{(m_{j}^{g})^{\frac{n-|r_{j}^{g}|}{2}}} \Bigg]. \\
\end{split}
\]
Further, by Lemmas \ref{VAR_JacobianUpBLemma} and \ref{VAR_emp_cov_lemma},
\[
\begin{split}
\frac{  r(G \mid Y)  }{  r(G_{o} \mid Y)  } & \le \Big(\frac{\pi}{n}\Big)^{\frac{|G| - |G_{o}|}{2}} \Big(\frac{n}{2}\Big)^{\frac{|G_{o}|}{2} + p} \frac{   E\Big( h\big(\balpha_{G}, \{\sigma_{j}\}\big) \Big) e^{\frac{1}{2}(1 - c)^{-2} \big(r_{\max}^{g} + (1 + c)^{2} \big) \frac{\|Y\|^{2}}{\sqrt{n}} - \frac{|G| + p}{2}}   }{   E\Big( h\big(\balpha_{G_{o}}, \{\sigma_{j}\}\big)|\widetilde{\D}'_{g_{o}}\widetilde{\D}_{g_{o}}|^{\frac{1}{2}} \Big)   } \\
& \hspace{.25in} \times n^{\frac{|G_{o}|-|G|}{2}}  e^{\frac{1}{2} \big( |G_{o}|[ \delta + \lambda_{\max}(\Gamma_{n}(0))] + |G|2\delta^{-1} \big) } \cdot \prod_{j=1}^{p}\Bigg[ \frac{(m_{j}^{g_{o}})^{\frac{n-|r_{j}^{g_{o}}|}{2}}}{(m_{j}^{g})^{\frac{n-|r_{j}^{g}|}{2}}} \Bigg] \\
\end{split}
\]
with probability exceeding $1 - 2V_{2}$, where $V_{2}$ is as in (\ref{VAR_V2}).  Then by Theorem \ref{VAR_TrueGraphLowBTheorem}, for the fixed $K_{3} \in (0,1)$,
\[
\begin{split}
\frac{  r(G \mid Y)  }{  r(G_{o} \mid Y)  } & \le \Big(\frac{\pi}{n}\Big)^{\frac{|G| - |G_{o}|}{2}} \Big(\frac{n}{2}\Big)^{\frac{|G_{o}|}{2} + p} \frac{   E\Big( h\big(\balpha_{G}, \{\sigma_{j}\}\big) \Big) e^{\frac{1}{2}(1 - c)^{-2} \big(r_{\max}^{g} + (1 + c)^{2} \big)\frac{\|Y\|^{2}}{\sqrt{n}} - \frac{|G| + p}{2}}   }{   (1-K_{3}) e^{\frac{|G_{o}|+p}{4}}   } \\
& \hspace{.25in} \times n^{\frac{|G_{o}|-|G|}{2}}  e^{\frac{1}{2} \big( |G_{o}|[ \delta + \lambda_{\max}(\Gamma_{n}(0))] + |G|2\delta^{-1} \big) } \cdot \prod_{j=1}^{p}\Bigg[ \frac{(m_{j}^{g_{o}})^{\frac{n-|r_{j}^{g_{o}}|}{2}}}{(m_{j}^{g})^{\frac{n-|r_{j}^{g}|}{2}}} \Bigg] \\
\end{split}
\]
with probability exceeding $1 - V_{1} - \widetilde{V}_{1} - 4V_{2} - 2e^{-\frac{np}{4}} - V_{3}$.  Gathering terms, for some positive constant $K_{2}$ (not depending on $n$ nor $p$),
\[
\begin{split}
\frac{  r(G \mid Y)  }{  r(G_{o} \mid Y)  } & \le E\Big( h\big(\balpha_{G}, \{\sigma_{j}\}\big) \Big)  e^{K_{2}\cdot \big(\frac{p\|Y\|^{2}}{\sqrt{n}} + p^{2}\log(n)\big)} \prod_{j=1}^{p}\Bigg[ \frac{(m_{j}^{g_{o}})^{\frac{n-|r_{j}^{g_{o}}|}{2}}}{(m_{j}^{g})^{\frac{n-|r_{j}^{g}|}{2}}} \Bigg] \\
\end{split}
\]
with probability exceeding $1 - V_{1} - \widetilde{V}_{1} - 4V_{2} - 2e^{-\frac{np}{4}} - V_{3}$.  At this point, $E\Big( h\big(\balpha_{G}, \{\sigma_{j}\}\big) \Big)$ can be bounded as in the following two cases.

{\noindent \bf Case 1: $G\subset G_{o}$ with $|G| \in \{1,\dots,|G_{o}|-1\}$.}
In this case, Theorem \ref{VAR_BigGraphUpBTheorem} does not apply, so since $h\big(\balpha_{G}, \{\sigma_{j}\}\big) \le 1$ uniformly,
\begin{equation}\label{VAR_K2}
\frac{  r(G \mid Y)  }{  r(G_{o} \mid Y)  } \le e^{K_{2}\cdot \big(\frac{p\|Y\|^{2}}{\sqrt{n}} + p^{2}\log(n)\big)} \prod_{j=1}^{p}\Bigg[ \frac{(m_{j}^{g_{o}})^{\frac{n-|r_{j}^{g_{o}}|}{2}}}{(m_{j}^{g})^{\frac{n-|r_{j}^{g}|}{2}}} \Bigg], 
\end{equation}
with probability exceeding $1 - V_{1} - \widetilde{V}_{1} - 4V_{2} - 2e^{-\frac{np}{4}} - V_{3}$.  Then by Condition \ref{VAR_condition5}, $\frac{  r(G \mid Y)  }{  r(G_{o} \mid Y)  } \overset{P_{x}}{\longrightarrow} 0$ as $n \to \infty$ or $n, p \to \infty$.

{\noindent \bf Case 2: $G\not\subseteq G_{o}$ and $|G| \in \{1,\dots,p^{2}\}$.}
By Lemma \ref{VAR_rss_ratio}, and for some positive constant $K_{1}$ (not depending on $n$ nor $p$),
\[
\begin{split}
\frac{  r(G \mid Y)  }{  r(G_{o} \mid Y)  } & \le E\Big( h\big(\balpha_{G}, \{\sigma_{j}\}\big) \Big)  e^{K_{2}\cdot \big(\frac{p\|Y\|^{2}}{\sqrt{n}} + p^{2}\log(n)\big)} \cdot \big((\sigma_{\max}^{0})^{2} 3 n\big)^{\frac{p^{2}}{2}}  e^{ (\sigma_{\max}^{0})^{2} p^{2} \sqrt{n} \frac{n}{2q}} \\
& \le E\Big( h\big(\balpha_{G}, \{\sigma_{j}\}\big) \Big)  e^{K_{1}\cdot \big(\frac{p\|Y\|^{2}}{\sqrt{n}} + p^{2}\log(n)\big)} \cdot e^{K_{1}\cdot \big( p^{2}\log(n) + \frac{n}{q} p^{2}\sqrt{n}\big) } \\
\end{split}
\]
with probability exceeding $1 - V_{1} - \widetilde{V}_{1} - 5V_{2} - 3e^{-\frac{np}{4}} - V_{3} - \frac{ 2 (\sigma_{\max}^{0})^{2} }{ \delta (1 - c^{2}) \sqrt{n} }$.  Therefore, by Theorem \ref{VAR_BigGraphUpBTheorem} and Condition \ref{VAR_condition4},
\begin{equation}\label{VAR_K1}
\begin{split}
\frac{  r(G \mid Y)  }{  r(G_{o} \mid Y)  } & \le \bigg(  e^{-\frac{\varepsilon}{9\Lambda_{g}}} + e^{-\big(\frac{d\cdot n^{\frac{\rho}{2}} p^{2}}{4\lambda_{\max}(\X\X'/n)} - \frac{np}{2}\big)} 2^{-\frac{|G|}{2}+1} \bigg) e^{K_{1}\cdot \big(\frac{p\|Y\|^{2}}{\sqrt{n}} + p^{2}\log(n) + \frac{n}{q} p^{2}\sqrt{n}\big)} \\
\end{split}
\end{equation}
with probability exceeding $1 - 2V_{1} - \widetilde{V}_{1} - 5V_{2} - 3e^{-\frac{np}{4}} - V_{3} - \frac{ 2 (\sigma_{\max}^{0})^{2} }{ \delta (1 - c^{2}) \sqrt{n} }$.  Thus, by Condition \ref{VAR_condition4}, $\frac{  r(G \mid Y)  }{  r(G_{o} \mid Y)  } \overset{P_{x}}{\longrightarrow} 0$ as $n \to \infty$ or $n, p \to \infty$.
$\hfill \blacksquare$


{\noindent \bf Proof of Corollary \ref{VAR_MainResult_corollary}.}
Omit case 1 in the proof of Theorem \ref{VAR_MainResult}.
$\hfill \blacksquare$


{\noindent \bf Proof of Corollary \ref{VAR_MainResult_strong_corollary}.}
Observe that 
\[
r(G_{o} \mid Y) = \frac{  r(G_{o} \mid Y)  }{  \sum_{j=1}^{p^{2}}\sum_{G:|G|=j} r(G \mid Y)  } = \frac{  1  }{  1 + \sum_{j=1}^{p^{2}}\sum_{G\ne G_{o}:|G|=j} \frac{  r(G \mid Y)  }{  r(G_{o} \mid Y)  }  }.
\]
Since $p$ is fixed Theorem \ref{VAR_MainResult} gives,
\[
\sum_{j=1}^{p^{2}}\sum_{G\ne G_{o}:|G|=j} \frac{  r(G \mid Y)  }{  r(G_{o} \mid Y)  } \overset{P_{x}}{\longrightarrow} 0
\]
as $n \to \infty$, which proves the desired result.
$\hfill \blacksquare$ 


\end{document}